\documentclass[fleqn,usenatbib]{mnras}

\usepackage[T1]{fontenc}

\DeclareRobustCommand{\VAN}[3]{#2}
\let\VANthebibliography\thebibliography
\def\thebibliography{\DeclareRobustCommand{\VAN}[3]{##3}\VANthebibliography}

\usepackage{graphicx}	
\usepackage{amsmath}	
\usepackage{amssymb}	
\usepackage{newtxtext,newtxmath}

\title[RGB photometric calibration of 15 million Gaia stars]
{RGB photometric calibration of 15 million \emph{Gaia} stars}

\author[N. Cardiel et al.]{%
Nicol\'{a}s Cardiel,$^{1,2}$\thanks{E-mail: cardiel@ucm.es (NC)}
Jaime Zamorano,$^{1,2}$
Josep Manel Carrasco,$^{6}$
Eduard Masana,$^{6}$
Salvador Bar\'{a},$^{3}$
\newauthor
Rafael Gonz\'{a}lez,$^{1}$
Jaime Izquierdo,$^{1}$
Sergio Pascual,$^{1,2}$
and Alejandro S\'{a}nchez de Miguel,$^{1,4,5}$
\\
$^{1}$Departamento de F\'{\i}sica de la Tierra y Astrof\'{\i}sica,
Fac.~CC.~F\'{\i}sicas, Universidad Complutense de Madrid, Plaza de las
Ciencias~1, E-28040, Spain\\
$^{2}$Instituto de F\'{\i}sica de Part\'{\i}culas y del Cosmos, IPARCOS,
Fac.~CC.~F\'{\i}sicas, Universidad Complutense de Madrid, Plaza de las
Ciencias~1, E-28040 Madrid, Spain\\
$^{3}$Departamento de F\'{\i}sica Aplicada, Universidade de Santiago de
Compostela, E-15782 Santiago de Compostela, Galicia, Spain\\
$^{4}$Environment and Sustainability Institute, University of Exeter, Penryn,
Cornwall TR10 9FE, UK\\
$^{5}$Instituto de Astrof\'{\i}sica de Andaluc\'{\i}a, Glorieta de la
Astronom\'{\i}a, s/n,C.P.18008 Granada, Spain\\
$^{6}$Departament F\'{\i}sica Qu\`{a}ntica i Astrofisica. Institut de
Ci\`{e}ncies del Cosmos (ICC-UB-IEEC), C Mart\'{\i} Franqu\`{e}s 1, Barcelona
08028, Spain
}

\date{Accepted XXX. Received YYY; in original form ZZZ}

\pubyear{2021}

\begin{document}
\label{firstpage}
\pagerange{\pageref{firstpage}--\pageref{lastpage}}
\maketitle

\begin{abstract}
Although a catalogue of synthetic RGB magnitudes, providing photometric data
for a sample of 1346 bright stars, has been recently published, its usefulness
is still limited due to the small number of reference stars available,
considering that they are distributed throughout the whole celestial sphere,
and the fact that they are restricted to Johnson~$V<6.6$~mag.
This work presents synthetic RGB magnitudes for $\sim15$~million stars brighter
than \emph{Gaia}~$G\!=\!18$~mag, making use of a calibration between
the RGB magnitudes of the reference bright star sample and the
corresponding high quality photometric $G$, $G_{\rm BP}$ and $G_{\rm RP}$
magnitudes provided by the \emph{Gaia} EDR3. The calibration has been
restricted to stars exhibiting \mbox{$-0.5 < G_{\rm BP}\!-\!G_{\rm RP} <
2.0$}~mag, and aims to predict RGB magnitudes within an error interval of $\pm
0.1$~mag. Since the reference bright star sample is dominated by nearby stars
with slightly undersolar metallicity, systematic variations in the predictions
are expected, as modelled with the help of stellar atmosphere models. These
deviations are constrained to the $\pm 0.1$~mag interval when applying the
calibration only to stars scarcely affected by interstellar extinction
and with metallicity compatible with the median value for the
bright star sample.
The
large number of \emph{Gaia} sources available in each region of the sky should
guarantee high-quality RGB photometric calibrations.
\end{abstract}





\section{Introduction}

Recently, \citet[hereafter C21]{2021MNRAS.504.3730C}\footnote{The main results
from that work are available online at \url{http://guaix.ucm.es/rgbphot},
and through VizieR at \url{http://vizier.u-strasbg.fr/viz-bin/VizieR?-source=J/MNRAS/504/3730}} have
established a standard RGB photometric system by setting its three basic
characteristics: i) a well-defined set of RGB spectral sensitivity curves,
determined from the median of a library of sensitivity curves
corresponding to 28~cameras analyzed by \citet{6475015}; ii) the use of
photon-based photometric magnitudes; and iii) the adoption of zero points
defined in the absolute (AB) scale. In addition, C21 have computed a catalogue
of synthetic RGB star magnitudes for 1346 bright stars belonging to the Bright
Star Catalogue \citep{1964cbs..book.....H}, using for that purpose historical
but very reliable 13-colour medium-narrow-band photometric data gathered by
\citet{1975RMxAA...1..299J}, \citet{1976RMxAA...1..327S} and
\citet{1997PASP..109..958B}, covering the \mbox{3370--11090}~\AA\ interval.
The RGB magnitudes in that work were determined from stellar
atmosphere models fitted to the 13-colour photometric data. The reliability of
the resulting spectral energy distributions was asserted through both the
comparison of synthetic Johnson $B$ and $V$ magnitudes with the corresponding
magnitudes in the Simbad
database\footnote{\url{http://simbad.u-strasbg.fr/simbad/}} (with a $3\sigma$
dispersion of 0.11 and 0.08~mag in the $B$ and $V$ band, respectively), and by
direct comparison with flux calibrated spectra from \citet{1987A&AS...69..465K}
(showing discrepancies below $\pm 0.05$~mag in the historical 13-colour
photometric bandpasses). Even though there are non-negligible variations of the
RGB spectral sensitivity curves between different cameras, C21 have also shown
that simple polynomial transformations can be employed to transform RGB
measurements performed with a typical camera to the mentioned standard system
(see their Fig.~18), facilitating the use of the proposed
system.

\begin{figure*}
\includegraphics[width=0.85\textwidth]{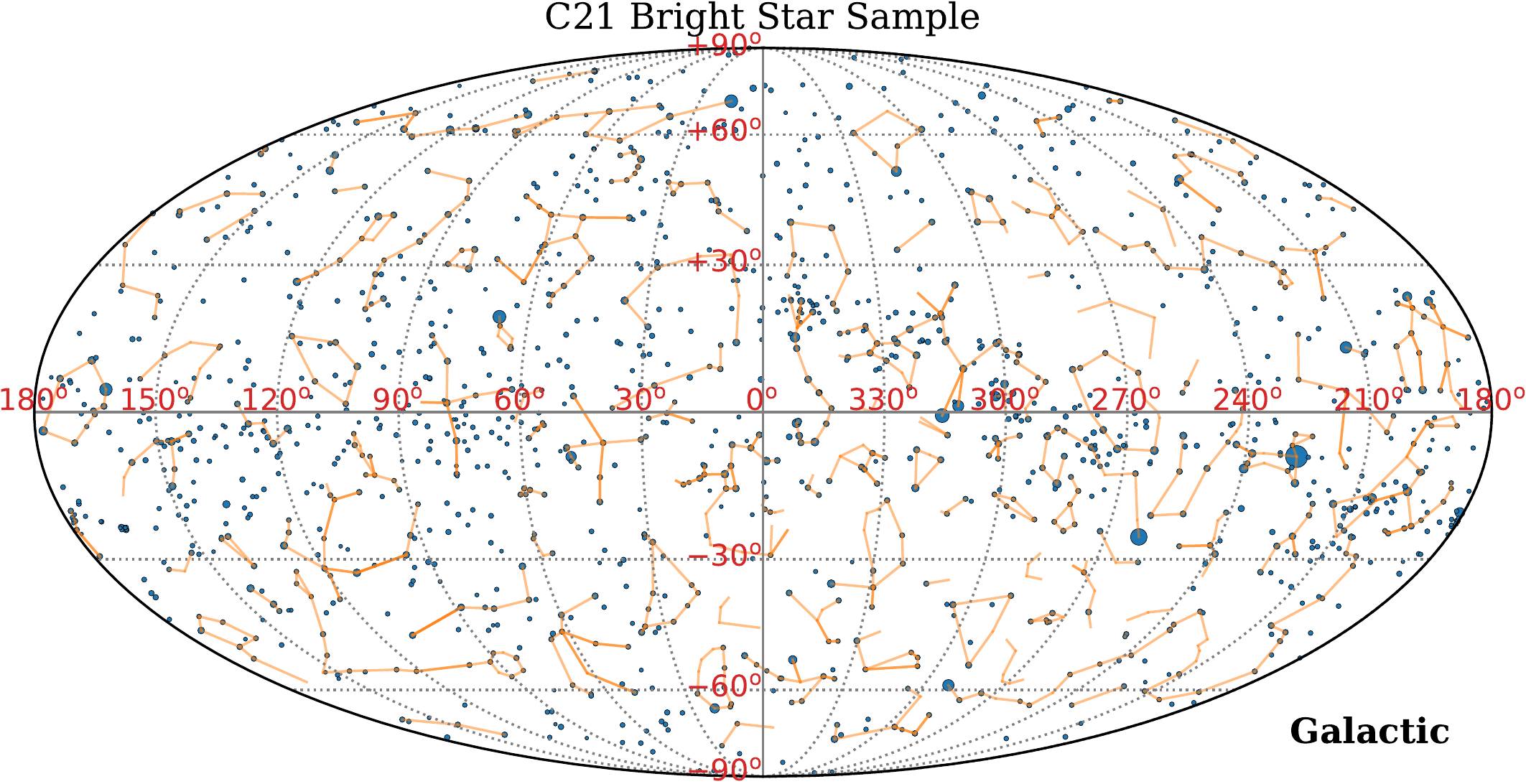}
\caption{Distribution in the celestial sphere (Galactic coordinates) of the C21
bright star sample, constituted by 1346 stars. Simple
constellation shapes \citep{2005Constellations}, displayed
with light orange lines, are also shown to facilitate the visual identification
of the stars. The size of each star symbol is inversely proportional to its
green RGB magnitude, as derived by C21.}
\label{fig:mollview_c21_bss}
\end{figure*}

Although the C21 catalogue of RGB magnitudes is suitable for calibration
purposes, it only contains a small number of stars (on average 1~star for each
30~square degrees, although they tend to concentrate towards
the Galactic plane; see Fig.~\ref{fig:mollview_c21_bss}).
Not only that, this catalogue is constituted by stars brighter
than Johnson~$V=6.6$~mag, with a magnitude distribution whose 16th, 50th and
84th percentiles are 3.3, 4.4 and 5.0~mag, respectively. This is specially
problematic when considering astronomical projects that could seriously benefit
from the exponential growth of the number of professional and
amateur astronomers equipped with commercial-grade RGB cameras, who can
potentially generate a huge amount of useful data in many astronomical fields
(see the Introduction section in C21 and references therein). 

The aim of this paper is to exploit the superb photometric data provided by the
\emph{Gaia} mission \citep{2016A&A...595A...1G}, through the third intermediate
\emph{Gaia} data release \citep[\emph{Gaia} EDR3]{2021A&A...649A...1G}, to
estimate RGB magnitudes from \emph{Gaia} $G$, $G_{\rm BP}$ and $G_{\rm RP}$
photometric data. For that purpose, we have derived simple transformations
between those magnitudes and the synthetic RGB photometry in the sample of
bright stars published by C21. In order to avoid confusion between \emph{Gaia}
$G$ magnitudes and the ones corresponding to the green RGB filter, from this
point we are using $G^{\rm Gaia}$, $G_{\rm BP}^{\rm Gaia}$ and $G_{\rm RP}^{\rm
Gaia}$ to indicate the use of the \emph{Gaia} magnitudes, and $B^{\rm rgb}$,
$G^{\rm rgb}$ and $R^{\rm rgb}$ when referring to the RGB photometric
measurements. 

Since the C21 star sample is constituted by bright and nearby stars of
metallicity slightly undersolar (as shown in Sect.~\ref{subsec:edr3_c21}), we
have also employed synthetic \emph{Gaia} and RGB magnitudes measured in
stellar atmosphere models to constrain the systematic uncertainties introduced
by the use of the mentioned transformations with stars of different metallicity
and also affected by interstellar extinction. In addition, we have applied the
fitted transformations to estimate RGB magnitudes for a sample of
$\sim15$~million stars extracted from the sample of
\citet{2019A&A...628A..94A}. These authors published improved photo-astrometric
distances, extinctions and astrophysical parameters for \emph{Gaia} DR2 stars
brighter than \mbox{$G=18$~mag}, using for that purpose the Bayesian tool
\texttt{StarHorse} \citep{2018MNRAS.476.2556Q}, complementing the \emph{Gaia}
measurements with Pan-STARRS1, 2MASS and AllWISE photometric data.  Here we
have restricted their stellar sample to minimize the impact of systematic
uncertainties, introduced by metallicity variations and interstellar
extinction, in the predictions of RGB magnitudes.

Synthetic magnitudes in this work have been determined using the Python package
{\sc synphot}
\citep{2018ascl.soft11001S}\footnote{\url{https://synphot.readthedocs.io/en/latest/}},
which facilitates the computation of photometric properties from user-defined
bandpasses and spectra (see Section 2 of C21 for additional computational
details).

The description of the \mbox{\emph{Gaia}--RGB} calibration is presented in
Section~\ref{sec:rgb_from_gaia}. Its application to the \texttt{StarHorse}
subsample of 15~million stars is described in Section~\ref{sec:starhorse},
while the final discussion and conclusions are summarized in
Section~\ref{sec:conclusions}. Appendix~\ref{ap:beyond_15M} describes how to
extend the estimation of RGB magnitudes beyond the 15~million
star sample, by using
ADQL queries to the \emph{Gaia} catalogue and an auxiliary Python package
specially written for that purpose.


\section{RGB calibration from Gaia EDR3 photometry}
\label{sec:rgb_from_gaia}

\begin{figure}
\includegraphics[width=\columnwidth]{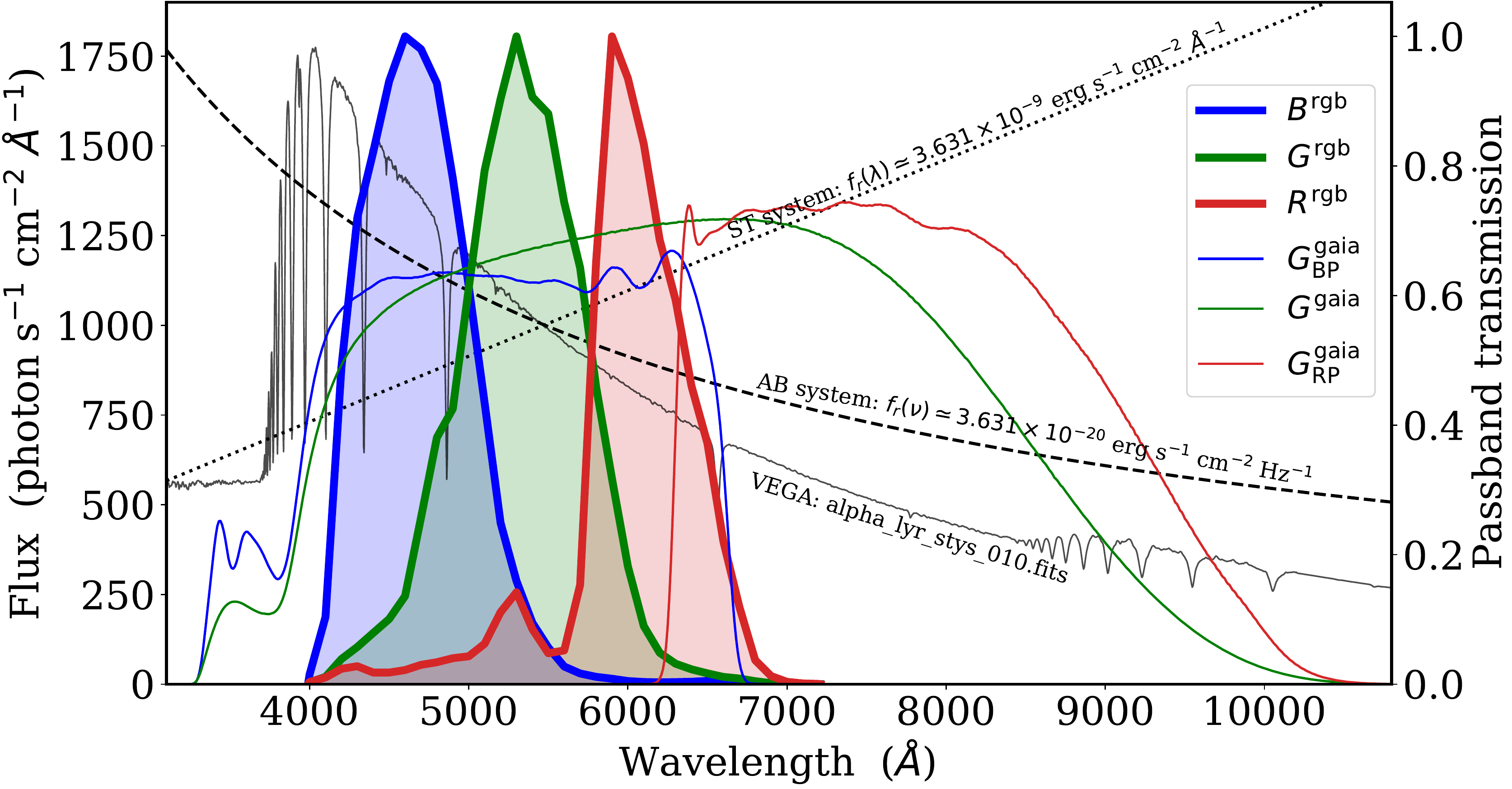}
\caption{Comparison of the transmissivity of the \emph{Gaia} EDR3 passbands
(thin blue, green and red lines), derived by \citet{2021A&A...649A...3R}, with
the standard RGB passbands defined in C21 (thick blue, green and red lines
encompassing shaded regions).
The flux density (in
\mbox{photon~s$^{-1}$~cm$^{-2}$~\AA$^{-1}$}) of the reference
spectra used to define magnitudes in the common AB, ST and Vega systems are 
shown as dashed, dotted and full black lines, respectively.}
\label{fig:vega_st_ab_comparison_gaia}
\end{figure}

The RGB sensitivity curves of the standard photometric system defined by C21
are encompassed by the transmissivity of the \emph{Gaia} EDR3 passbands derived
by \citet{2021A&A...649A...3R}\footnote{Available at
\url{https://www.cosmos.esa.int/web/gaia/edr3-passbands}} (see
Fig.~\ref{fig:vega_st_ab_comparison_gaia}). In particular, the
three RGB transmissivity curves cover a similar wavelength range as
$G_{\rm BP}^{\rm Gaia}$, and approximately half of the range spanned by $G^{\rm
Gaia}$. The fact that $G_{\rm RP}^{\rm Gaia}$ covers an additional range
towards longer wavelengths, makes the colour \mbox{$G_{\rm BP}^{\rm Gaia}
- G_{\rm RP}^{\rm Gaia}$} a good proxy to estimate variations in the spectral
energy distribution covered in the visible range. For that reason,
it is expected that a good calibration of RGB magnitudes can
be derived from the accurate \emph{Gaia} data. In this section we describe the
procedure followed to achieve this task, starting by collecting the \emph{Gaia}
magnitudes available for the C21 bright star sample, creating colour-colour
diagrams involving \emph{Gaia} and RGB magnitudes, and finding a simple
mathematical relationship between both magnitude sets.

\subsection{\emph{Gaia} EDR3 data for the C21 bright star sample}
\label{subsec:edr3_c21}

\begin{figure}
\includegraphics[width=\columnwidth]{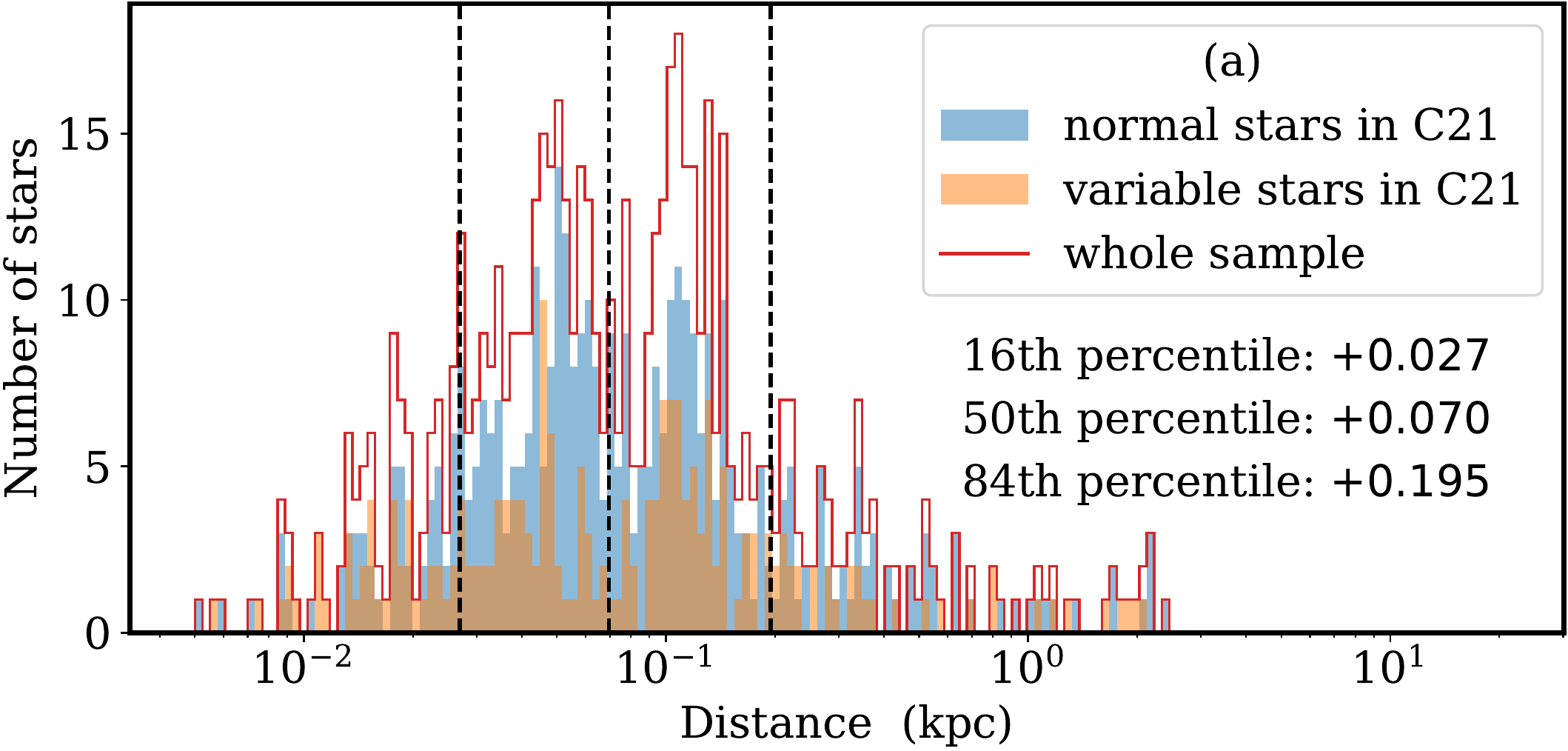}
\\[4pt]
\includegraphics[width=\columnwidth]{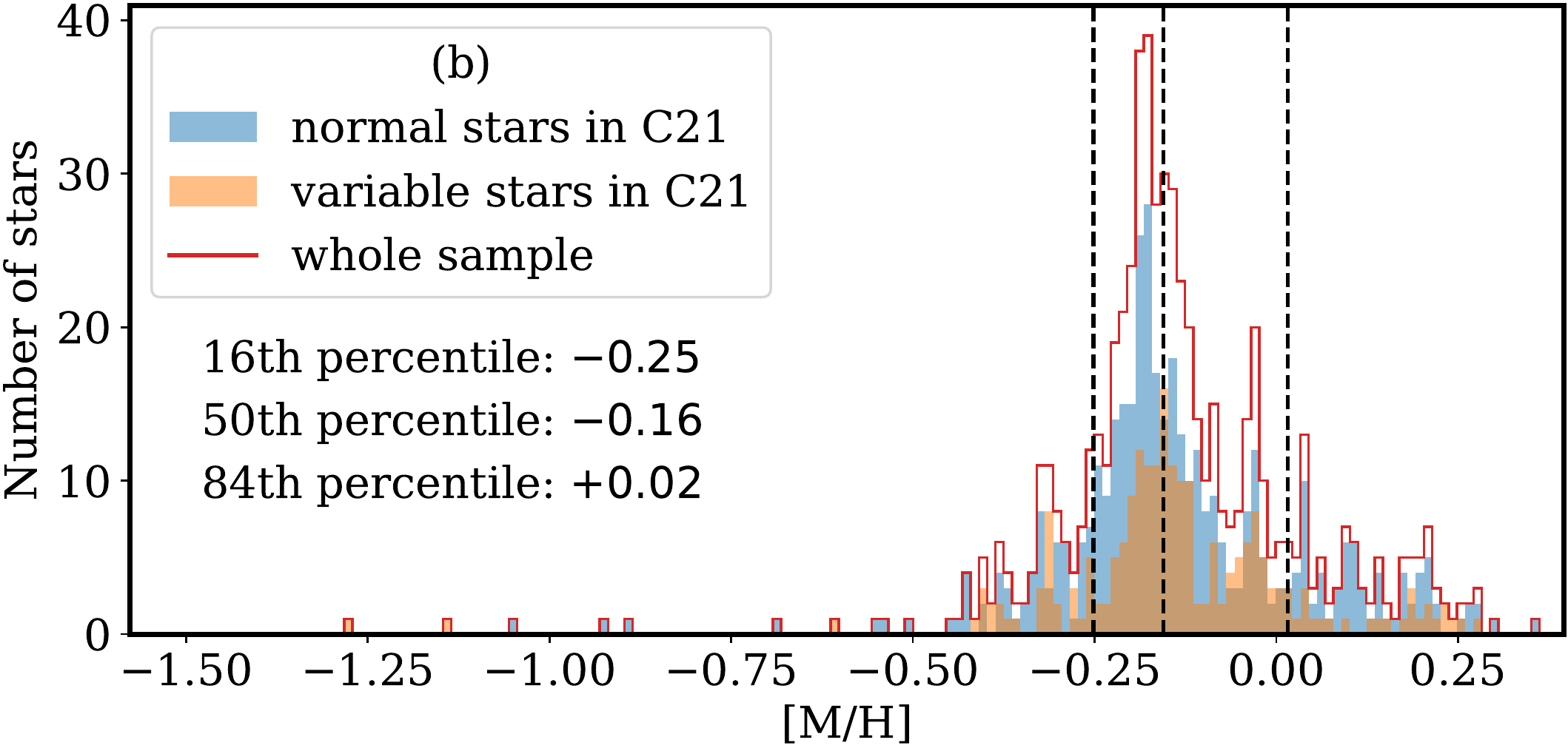}
\\[4pt]
\includegraphics[width=\columnwidth]{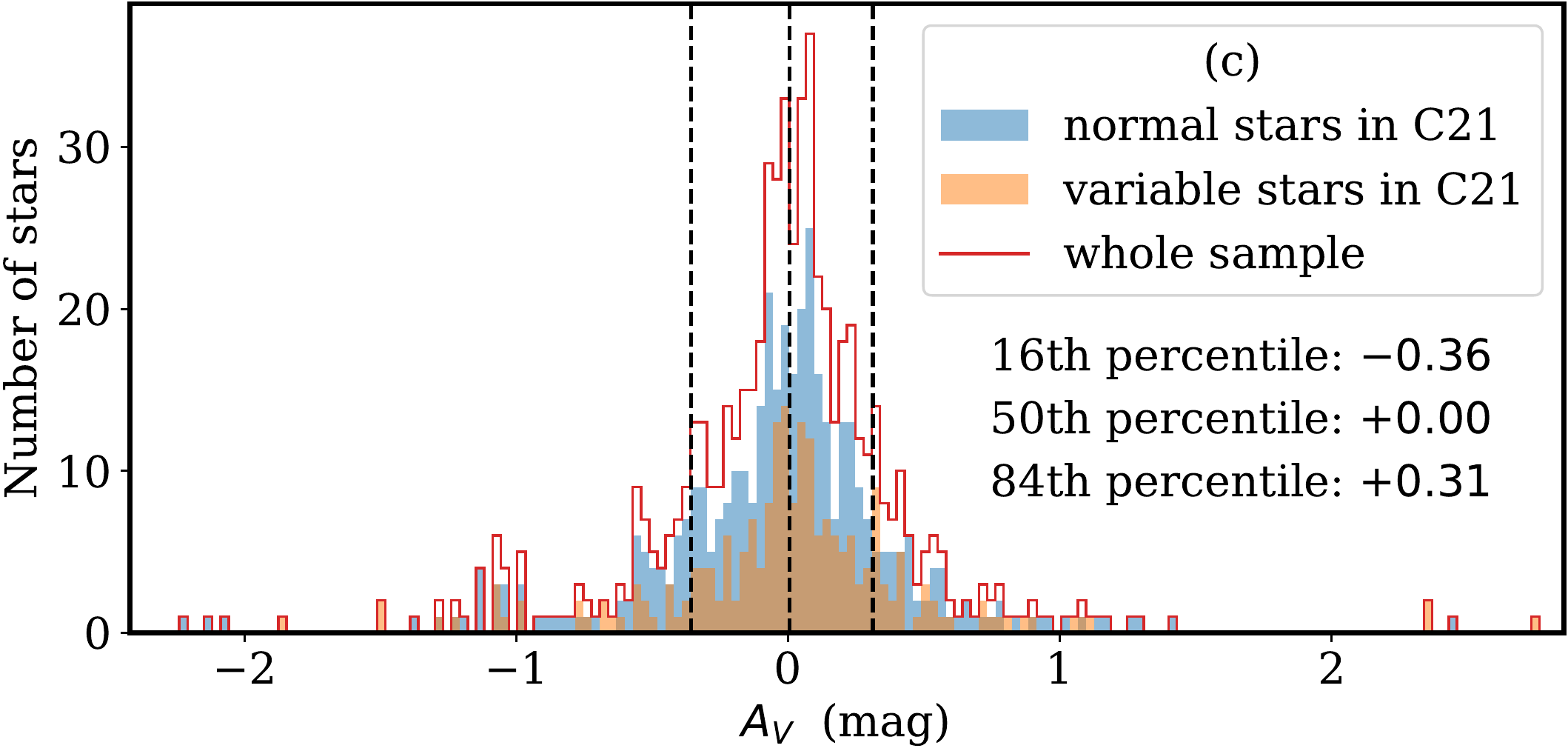}
\\[4pt]
\includegraphics[width=\columnwidth]{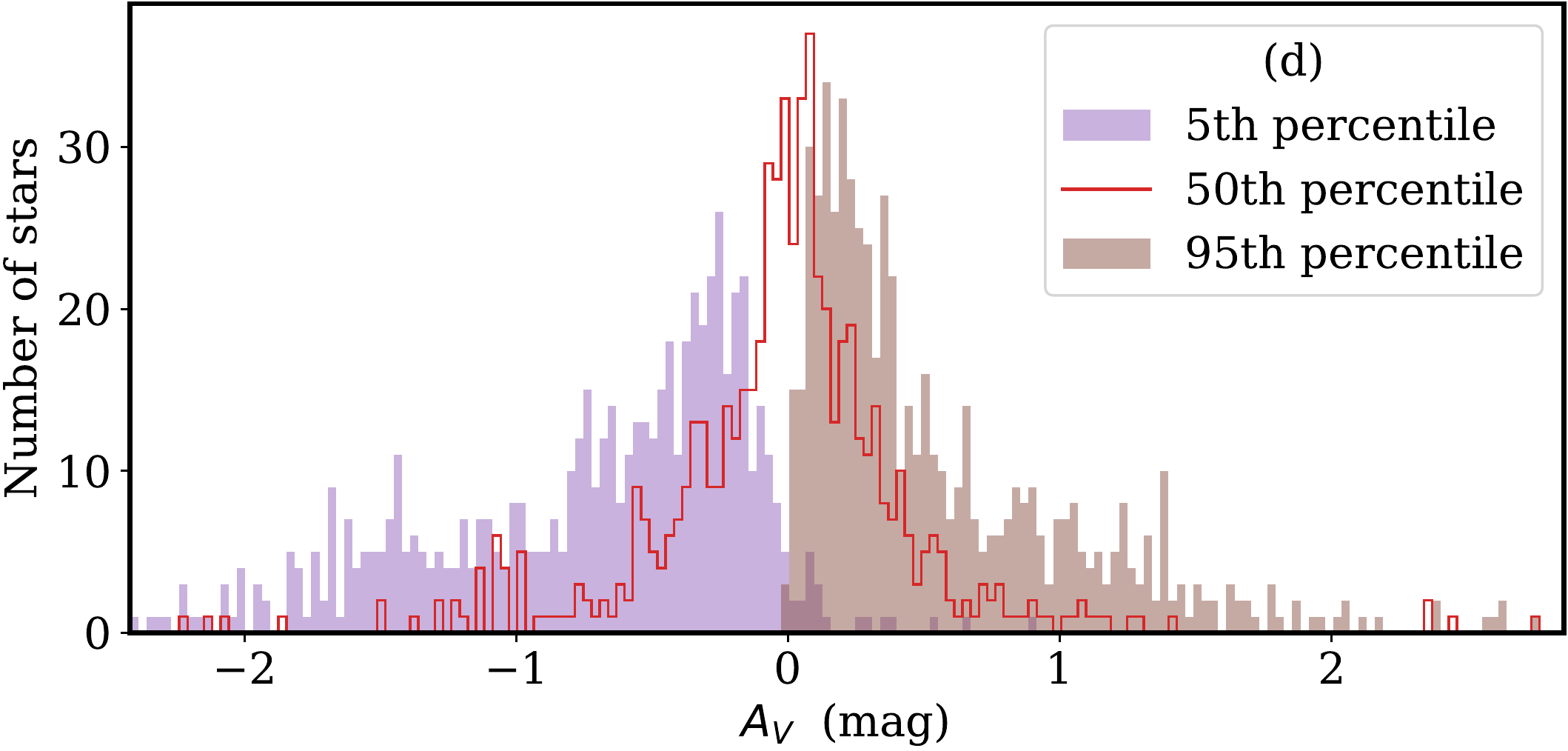}
\caption{\emph{Panels (a)--(c):} histograms displaying the distributions in
distance [panel~(a)], metallicity [panel~(b)] and $A_{V}$ extinction
[panel~(c)] for the C21 bright star subsample in common with
\citet{2019A&A...628A..94A}. These three parameters are the ones derived by
these authors using \texttt{StarHorse}.  Although initially 785~stars had
parameter estimates, 172~stars were discarded by imposing the quality flag
\texttt{SH\_OUTFLAG='00000'}, which guarantees the
statistical reliability of the \texttt{StarHorse}
determinations.  The histograms have been computed separately for the
non-variable (in blue) and variable (orange) stars, with the red line
delineating the coadded histogram. The three vertical dashed lines indicate the
16th, 50th and 84th percentiles (from left to right, respectively) of the whole
displayed sample, whose values are provided under the plot legend.
\emph{Panel~(d):} histogram displaying the 5th, 50th and 95th
percentile of the posterior probability distribution for the \texttt{StarHorse}
$A_V$ estimates.  Note that the 50th percentile histogram is the same displayed
in panel~(c) for the whole sample with reliable parameters, whereas the 95th
percentile is greater than zero for all the stars with
\texttt{SH\_OUTFLAG='00000'}. This indicates that the negative $A_V$ values
correspond to stars whose Bayesian posterior probability distribution is not
statistically incompatible with a null extinction.}
\label{fig:hist_C21_SH}
\end{figure}

As the first step for this work, we retrieved the 
$G^{\rm Gaia}$, $G_{\rm BP}^{\rm Gaia}$ and $G_{\rm RP}^{\rm Gaia}$ magnitudes 
for the bright star sample of C21, provided by the \emph{Gaia} EDR3
\citep{2021A&A...649A...1G} Archive at the
European Space Agency\footnote{\url{https://gea.esac.esa.int/archive/}}. This
process was carried out in the following way: the \emph{Gaia}~DR2 identifier of
each star was initially obtained through the Simbad
database, starting from the
initial HR~number of the star in the Bright Star Catalogue
\citep{1964cbs..book.....H}. Since it is not guaranteed that the same
astronomical source will always have the same source identifier in the
different \emph{Gaia} Data Releases, for each \texttt{source\_id} in DR2 we
searched for proximal source(s) in the auxiliary table
\texttt{gaiaedr3.dr2\_neighbourhood}, keeping the most likely crossmatch taking
into account the parameters \texttt{angular\_distance} and
\texttt{magnitude\_difference} when more that one
possible counterpart appeared. It is important to note that not all the bright
stars in the C21 sample appear in the \emph{Gaia} database because of its
level of incompleteness at the bright end
\citep{2021A&A...649A...1G}. In addition, and having in mind
that our main goal is to derive a photometric transformation, we
imposed the signal-to-noise ratio for the flux measured by \emph{Gaia} in each
passband to be high enough to guarantee a maximum uncertainty of 0.01~mag for
both $G^{\rm Gaia}$ magnitudes and the \mbox{$G_{\rm BP}^{\rm Gaia}\!-\!G_{\rm
RP}^{\rm Gaia}$} colours. At the end of this process, the initial sample of
1346 bright stars was finally reduced to a subsample of 888~objects, 320 of
them flagged as variable stars in Simbad, and 568 with no indication of
variability. The histograms displayed in Fig.~\ref{fig:hist_C21_SH} confirm
that the bulk of this bright star C21 subsample is dominated by nearby stars,
with metallicity slightly undersolar and scarcely affected by interstellar
extinction. Although the large negative extinction
estimates displayed in Fig.~\ref{fig:hist_C21_SH}(c) may seem alarming, it is
important to realize that they simply correspond to the median values of the
$A_V$ Bayesian posterior probability distribution derived by
\citet{2019A&A...628A..94A} for each star, and thus, a too simplistic
reduction of whole probability distributions into single numbers. The 5th and
95th percentile $A_V$ values displayed in  Fig.~\ref{fig:hist_C21_SH}(d)
illustrate that the credible intervals for those stars are compatible with
$A_V=0$.

\subsection{Colour--colour diagram relating RGB and \emph{Gaia}}

\begin{figure*}
\includegraphics[width=\columnwidth]{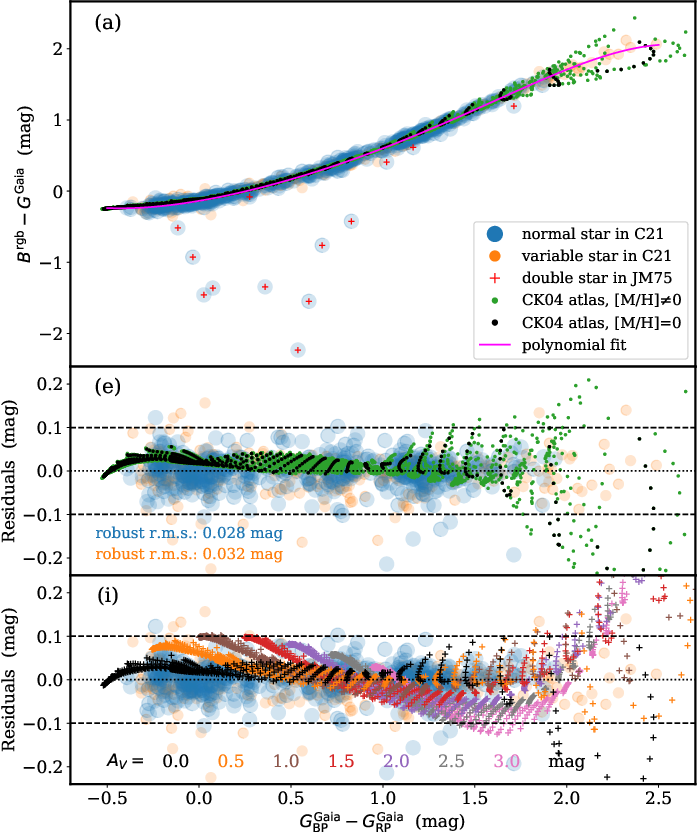}
\hfill
\includegraphics[width=\columnwidth]{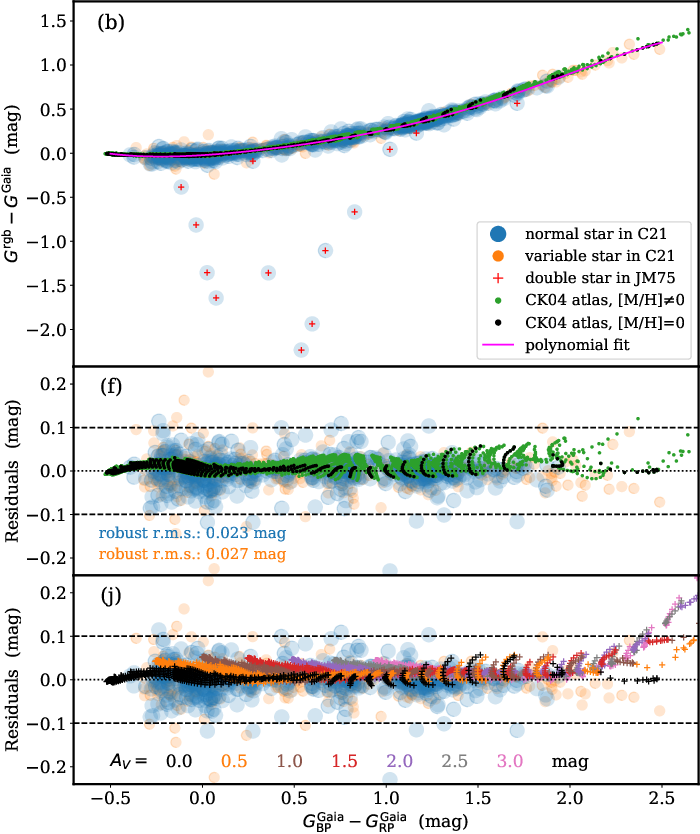}

\vskip 3mm

\includegraphics[width=\columnwidth]{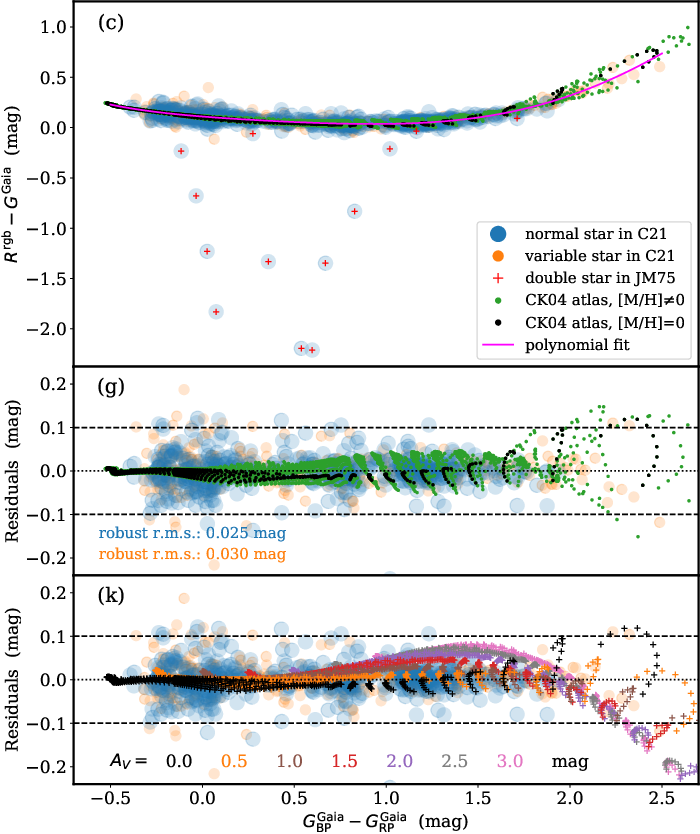}
\hfill
\includegraphics[width=\columnwidth]{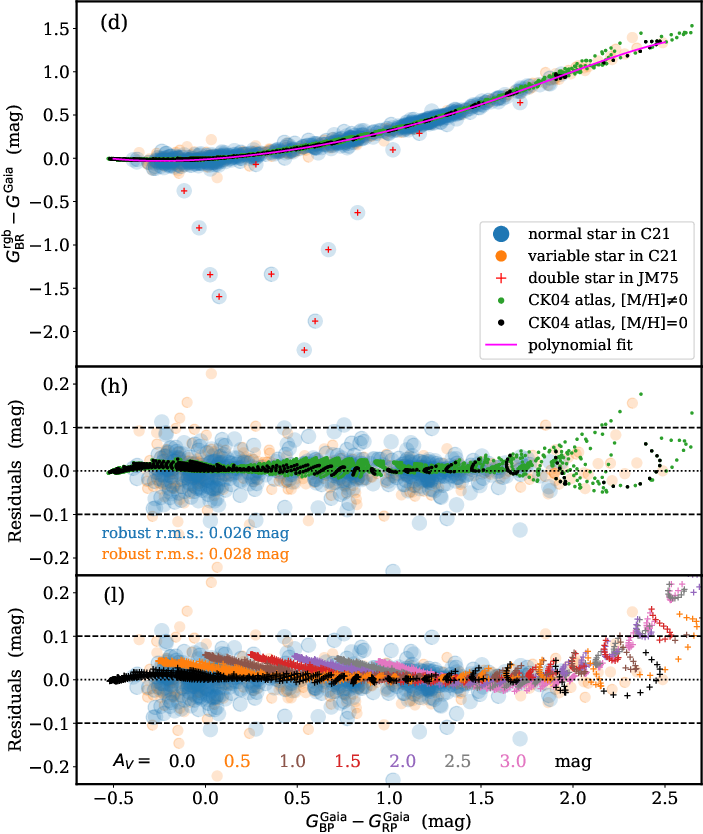}
\caption{\emph{Panels~(a), (b), (c) and (d):}\/ colour-colour diagrams
representing the differences between each RGB magnitude (and
$G^{\rm rgb}_{\rm BR}$) with $G^{\rm Gaia}$ as a function of
the \emph{Gaia} colour \,$G_{\rm BP}^{\rm Gaia}\!-\!G_{\rm RP}^{\rm Gaia}$. The
bright star sample from C21 is plotted with filled blue (non-variable stars)
and orange (stars flagged as variable in Simbad) circles. Clear outliers,
overplotted with a small red cross, correspond to double stars in
\citet{1975RMxAA...1..299J} (JM75, see text). The
predictions of CK04 models are also shown with small black (solar metallicity)
and green (non-solar metallicity) circles. The calibrations, robust fits to
5-degree polynomials, are shown as magenta lines, with the corresponding
residuals shown (twice) in the underneath panels. 
\emph{Panels~(e), (f), (g) and (h)}: the
r.m.s.\ of the residuals is provided separately for the non-variable (blue) and
variable (orange) stars. The horizontal dashed lines encompass the $\pm
0.1$~mag interval.  \emph{Panels~(i), (j), (k) and (l)}: the coloured crosses
indicate the impact of interstellar reddening, as parametrized by the indicated
$A_{V}$ values, for CK04 models of solar metallicity (non-solar predictions
have not been used in this case to avoid cluttering).}
\label{fig:predict_RGB_from_gaia}
\end{figure*}

We represent in the top panels of Fig.~\ref{fig:predict_RGB_from_gaia} the
colour--colour diagrams built using the difference between each RGB magnitude
(from the C21 catalogue) and $G^{\rm Gaia}$, as a function of the \emph{Gaia}
colour \mbox{$G_{\rm BP}^{\rm Gaia}\!-\!G_{\rm RP}^{\rm Gaia}$}.  We have also
included in this comparison an estimate of the magnitude in the green RGB
filter computed simply as the arithmetic mean of the two magnitudes in the
neighbouring filters,~i.e.,
\begin{align}
G^{\rm rgb}_{\rm BR} = \frac{B^{\rm rgb}+R^{\rm rgb}}{2}.
\end{align}
The definition of this additional magnitude can be particularly
useful for observations performed with cameras equipped with Bayer-like colour
filter systems\footnote{See U.S.\ Patent No. 3,971,065,
available at
\url{https://patents.google.com/patent/US3971065}}, in which
an array of luminance- and chrominance-sensitivity elements (the green pixels
on the one hand, and the blue and red pixels, on the other) are superposed in
registration with an imaging array. Since the spatial sampling of each pixel is
different, the computation of an averaged (blue+red) magnitude can facilitate
the estimation of an independent green magnitude which, in addition, can help
to perform image interpolations through the comparison of the demosaiced RGB
channels.

Non-variable stars in the colour-colour diagrams
shown in Fig.~\ref{fig:predict_RGB_from_gaia} are represented with big
blue filled circles, whereas variable objects are plotted with smaller orange
filled circles. It is clear that the subsample of C21 bright stars with
\emph{Gaia} data exhibits well-defined sequences in these diagrams. A few
objects, outliers in the four panels and marked with an additional red cross,
correspond to stars flagged as double in the initial photometric measurements
by \citet{1975RMxAA...1..299J} (see also Table~3 in C21) and were not used in
the subsequent work. We have also overplotted the expected location of
synthetic colour estimates computed from the stellar atmosphere models by
\citet[hereafter CK04]{2003IAUS..210P.A20C}, which are precomputed for
abundances ${\rm [M/H]}\!=\!-2.5$, $-2.0$, $-1.5$, $-1.0$, $-0.5$, $0.0$,
$+0.2$ and~$+0.05$, effective temperatures ranging from 3500 to 50000~K, and
$\log g$ (surface stellar gravity, with $g$ in cm~s$^{-2}$) from 0.0 to
5.0~dex. Note that these are the same models employed by C21 to derive their
RGB photometric database. In particular, we have used two different symbol
colours to represent the predictions for solar metallicity stars (small black
filled circles) and for non-solar metallicity objects (small green filled
circles). It is important to highlight here that the RGB photometric system is
defined in the absolute (AB) scale, whereas the \emph{Gaia} magnitudes are
provided in the Vega system. The latter have been computed for the synthetic
magnitudes of the stellar models using the Vega spectrum
\texttt{alpha\_lyr\_stys\_010.fits}, available at the CALSPEC
database\footnote{\url{https://www.stsci.edu/hst/instrumentation/reference-data-for-calibration-and-tools/astronomical-catalogs/calspec}}
\citep{2014PASP..126..711B}.

\subsection{The \emph{Gaia}-RGB calibration}

\begin{table}
\centering
\caption{Coefficients of the polynomials $f_{B}$, $f_{G}$, $f_{R}$ and
$f_{G_{\rm BR}}$ defined in Eqs.~(\ref{eq:polyB})--(\ref{eq:polyX}), and
corresponding to the fits displayed with magenta lines in panels~(a), (b), (c)
and~(d) of Fig.~\ref{fig:predict_RGB_from_gaia}. Each polynomial must be
evaluated as \, $f_{X} = \sum_{i=0}^5 a_i \, (G_{\rm BP}^{\rm Gaia}\!-\!G_{\rm
RP}^{\rm Gaia})^{i}$, with $X=B,\, G,\, R$
and~$G_{\rm BR}$.}
\label{tab:polynomial_fits}
\begin{tabular}{c@{\;\;\;}c@{\;\;\;}c@{\;\;\;}c@{\;\;\;}c}
\hline
Coef.  & $f_{B}$ & $f_{G}$ & $f_{R}$ & $f_{G_{\rm BR}}$\\
\hline
$a_0$  & $ -0.13748689$ & $ -0.02330159$ & $ +0.10979647$ & $ -0.01252185$ \\
$a_1$  & $ +0.44265552$ & $ +0.12884074$ & $ -0.14579334$ & $ +0.13983574$ \\
$a_2$  & $ +0.37878846$ & $ +0.22149167$ & $ +0.10747392$ & $ +0.23688188$ \\
$a_3$  & $ -0.14923841$ & $ -0.14550480$ & $ -0.10635920$ & $ -0.10175532$ \\
$a_4$  & $ +0.09172474$ & $ +0.10635149$ & $ +0.08494556$ & $ +0.07401939$ \\
$a_5$  & $ -0.02594726$ & $ -0.02363990$ & $ -0.01368962$ & $ -0.01821150$ \\
\hline
\end{tabular}
\end{table}

The sequences displayed in the top panels of
Fig.~\ref{fig:predict_RGB_from_gaia} have been iteratively fitted (rejecting
outliers using a sigma-clipping algorithm) to the previously described
subsample of 888 bright stars, using fifth-order polynomials\footnote{The final
polynomial degree was determined by using an orthogonal polynomial regression
with the help of the software package~R \citep{Rcitation}. This facilitated
the determination of the maximum polynomial degree that led to polynomial
coefficients that were statistically significant.}.
Since the stars included in this subsample
flagged as variable objects already passed the filtering process described in
C21 when comparing synthetic Johnson $B$ and $V$ magnitudes with tabulated
measurements in the Simbad database, and considering that they do not show a
different behaviour to that exhibited by non-variable stars, we kept them in
the initial set to be fitted, knowing that the sigma-clipping algorithm would
get rid off the deviant cases. With the aim of constraining the fits at the
extremes of the \mbox{$G_{\rm BP}^{\rm Gaia}\!-\!G_{\rm RP}^{\rm Gaia}$} colour
interval, the fitted data were complemented by including the predictions of the
CK04 models of solar metallicity for \mbox{$-0.5 < G_{\rm BP}^{\rm
Gaia}\!-\!G_{\rm RP}^{\rm Gaia} < -0.4$ mag} and for \mbox{$2.0 < G_{\rm
BP}^{\rm Gaia}\!-\!G_{\rm RP}^{\rm Gaia} < 2.5$ mag} (no additional model data
were employed except for those two small colour ranges at the borders). The
fitted relationships are displayed in the top panels of
Fig.~\ref{fig:predict_RGB_from_gaia} with a continuous magenta line, and the
resulting residuals are plotted (twice) in the central and bottom panels. These
polynomial fits provide the sought transformation that allows estimating RGB
magnitudes from the \emph{Gaia} magnitudes using
\begin{align}
\label{eq:polyB}
B^{\rm rgb} = & \mbox{}\;\; G^{\rm Gaia} + 
f_{B}(G_{\rm BP}^{\rm Gaia}\! - \!G_{\rm RP}^{\rm Gaia}), \\
\label{eq:polyG}
G^{\rm rgb} = & \mbox{}\;\; G^{\rm Gaia} + 
f_{G}(G_{\rm BP}^{\rm Gaia}\! - \!G_{\rm RP}^{\rm Gaia}), \\
\label{eq:polyR}
R^{\rm rgb} = & \mbox{}\;\; G^{\rm Gaia} + 
f_{R}(G_{\rm BP}^{\rm Gaia}\! - \!G_{\rm RP}^{\rm Gaia}), \\
\label{eq:polyX}
G^{\rm rgb}_{\rm BR} = & \mbox{}\;\; G^{\rm Gaia} + 
f_{G_{\rm BR}}(G_{\rm BP}^{\rm Gaia}\! - \!G_{\rm RP}^{\rm Gaia}),
\end{align}
where $f_{B}$, $f_{G}$, $f_{R}$ and $f_{G_{\rm BR}}$ are the fifth-order
polynomials with independent variable \mbox{$G_{\rm BP}^{\rm Gaia}\!-\!G_{\rm
RP}^{\rm Gaia}$}, whose coefficients are given in
Table~\ref{tab:polynomial_fits}.

The residuals are displayed in the central panels of
Fig.~\ref{fig:predict_RGB_from_gaia}, using the same symbols employed in the
top panels. The robust standard deviation of these residuals\footnote{Computed
as $\sigma_{\rm robust}=0.7413 (q_{75} - q_{25})$, with $q_{25}$ and $q_{75}$
the 25th and 75th percentiles, respectively \citep[see e.g.,][]{Ivezic2020}} is
also provided within each panel, computed separately for the 568 non-variable
stars (blue colour) and the 320 variable stars (orange colour), being the latter
slightly larger in the four panels. In all cases, the $\pm 3\sigma_{\rm
robust}$ dispersion is constrained within the $\pm 0.1$~mag interval (displayed
with the horizontal dashed lines). The same panels also display the residuals
corresponding to the synthetic magnitudes derived from the CK04 models: the
predictions for solar metallicity stars (small black filled circles) are well
reproduced by the previous fitted relationships, except for the reddest stars,
where the dispersion increases. The models with metallicity different from
solar (small green filled circles) exhibit a wider scatter.
It is interesting to note that the model predictions
interpolated for ${\rm [M/H]}\!=\!-0.16$, the median value in
Fig.~\ref{fig:hist_C21_SH}(b), are
not very different from those for solar metallicity. In particular, the
comparison between all the models predictions leads to
\begin{align}
\Delta(G_{\rm BP}^{\rm Gaia}-G_{\rm RP}^{\rm Gaia}) &= 0.004\pm0.014\;
{\rm mag},\\
\Delta(B^{\rm rgb}-G^{\rm Gaia}) &= 0.004\pm0.008\; {\rm mag},\\
\Delta(G^{\rm rgb}-G^{\rm Gaia}) &= 0.002\pm0.009\; {\rm mag},\\
\Delta(R^{\rm rgb}-G^{\rm Gaia}) &= 0.002\pm0.009\; {\rm mag,\; and}\\
\Delta(G_{\rm BR}^{\rm rgb}-G^{\rm Gaia}) &= 0.002\pm0.010\; {\rm mag},
\end{align}
where each $\Delta$ value corresponds to the mean (and associated standard
deviation) colour difference between the ${\rm [M/H]}\!=\!0.0$ and the ${\rm
[M/H]}\!=\!-0.16$ predictions.
Interestingly, all
the models fit within the $\pm 0.1$~mag interval for $G^{\rm rgb}$ and $G^{\rm
rgb}_{\rm BR}$,
whereas the same is true for $B^{\rm rgb}$ and $R^{\rm rgb}$ when restricting
to the \mbox{$-0.5 < G_{\rm BP}^{\rm Gaia}\!-\!G_{\rm RP}^{\rm Gaia} < 2.0$
mag} range.

The residuals of the fitted data are displayed again in the bottom panels of
Fig.~\ref{fig:predict_RGB_from_gaia}, but in this case the overplotted model
predictions represent the residuals of the CK04 models after reddening them
employing the extinction law of \citet{1989ApJ...345..245C} \citep[updated
by][]{1994ApJ...422..158O}, with a relative extinction parameter $R_{V}=3.1$
and variable extinctions $A_{V}$ ranging from~0.5 to 3.0~mag (crosses of
different colours). Interestingly, the extinction reddens the data basically
along the sequences already displayed in the top panels, and therefore the
residuals in the bottom panels remain constrained within the $\pm 0.1$~mag
interval. The effect is larger in $B^{\rm rgb}$ and $R^{\rm rgb}$ than in
$G^{\rm rgb}$ and $G^{\rm rgb}_{\rm BR}$, with a systematic variation as a
function of the $G_{\rm BP}^{\rm Gaia}\!-\!G_{\rm RP}^{\rm Gaia}$ colour.

\section{Predicting RGB magnitudes for 15 million stars}
\label{sec:starhorse}

\begin{table*}
\centering
\caption{First ten rows of the table with the \mbox{14\,854\,959} stars with RGB
predictions. The full table is electronically available at
\url{http://guaix.fis.ucm.es/~ncl/rgbphot/gaia}. Column description: (1)~source
identifier in the \emph{Gaia}~EDR3; (2)--(3) star coordinates (J2000), as
provided by the EDR3; (4)--(7)~RGB magnitudes computed using the
transformations given by Eqs.~(\ref{eq:polyB})--(\ref{eq:polyX});
(8)--(10)~\emph{Gaia} magnitudes in the EDR3; (11)--(13)~50th percentile
(median) of the interstellar extinction $A_{V}$ (mag), [M/H] and distance
(kpc), computed by \citet{2019A&A...628A..94A} using \texttt{StarHorse}.}
\label{tab:rgb_results}
\begin{tabular}{rr@{\;\;}rc@{\;\;}c@{\;\;}c@{\;\;}cc@{\;\;}c@{\;\;}cc@{\;\;}c@{\;\;}c}
\hline
\multicolumn{1}{c}{(1)} & 
\multicolumn{1}{c}{(2)} & 
\multicolumn{1}{c}{(3)} & 
\multicolumn{1}{c}{(4)} & 
\multicolumn{1}{c}{(5)} & 
\multicolumn{1}{c}{(6)} & 
\multicolumn{1}{c}{(7)} & 
\multicolumn{1}{c}{(8)} & 
\multicolumn{1}{c}{(9)} & 
\multicolumn{1}{c}{(10)} & 
\multicolumn{1}{c}{(11)} & 
\multicolumn{1}{c}{(12)} & 
\multicolumn{1}{c}{(13)} \\
\multicolumn{1}{c}{EDR3 source\_id} & 
\multicolumn{1}{c}{RA (\textdegree)} & 
\multicolumn{1}{c}{DEC (\textdegree)} & 
\multicolumn{1}{c}{$B^{\rm rgb}$} & 
\multicolumn{1}{c}{$G^{\rm rgb}$} & 
\multicolumn{1}{c}{$R^{\rm rgb}$} & 
\multicolumn{1}{c}{$G^{\rm rgb}_{\rm BR}$} & 
\multicolumn{1}{c}{$G^{\rm Gaia}$} & 
\multicolumn{1}{c}{$G^{\rm Gaia}_{\rm BP}$} & 
\multicolumn{1}{c}{$G^{\rm Gaia}_{\rm RP}$} & 
\multicolumn{1}{c}{av50} & 
\multicolumn{1}{c}{met50} & 
\multicolumn{1}{c}{dist50} \\
\hline
2875513285079465984 & 0.000095821 & $+34.987630914$ & 17.03 & 16.46 & 16.06 & 16.55 & 15.9741 & 16.6310 & 15.2054 & $+0.120$ & $-0.188$ & $+0.612$ \\
 393915403758148352 & 0.000114124 & $+50.277031947$ & 18.58 & 18.14 & 17.84 & 18.21 & 17.7949 & 18.3144 & 17.1386 & $+0.158$ & $-0.005$ & $+1.818$ \\
4702040864637495296 & 0.000141937 & $-71.574692515$ & 18.15 & 17.76 & 17.49 & 17.82 & 17.4528 & 17.9223 & 16.8279 & $-0.239$ & $-0.111$ & $+1.720$ \\
 384492417301984256 & 0.000168528 & $+43.261207984$ & 16.71 & 16.43 & 16.25 & 16.48 & 16.2111 & 16.5778 & 15.6839 & $+0.122$ & $+0.027$ & $+1.952$ \\
2746773530168227328 & 0.000178628 & $+08.371651685$ & 12.85 & 12.40 & 12.09 & 12.47 & 12.0394 & 12.5638 & 11.3628 & $+0.197$ & $-0.048$ & $+0.143$ \\
2855280484422222336 & 0.000262145 & $+29.896074086$ & 18.08 & 17.45 & 16.99 & 17.54 & 16.8718 & 17.6054 & 16.0451 & $-0.129$ & $-0.019$ & $+0.771$ \\
2875090969535427072 & 0.000301072 & $+33.520870876$ & 16.81 & 16.53 & 16.34 & 16.57 & 16.3041 & 16.6743 & 15.7705 & $+0.209$ & $-0.224$ & $+1.637$ \\
2443095153084654080 & 0.000434825 & $-05.494409354$ & 09.63 & 09.21 & 08.92 & 09.27 & 08.8720 & 09.3678 & 08.2151 & $+0.139$ & $-0.316$ & $+0.518$ \\
4923847544332011520 & 0.000435338 & $-54.929637980$ & 14.29 & 14.10 & 13.99 & 14.14 & 13.9466 & 14.2257 & 13.5046 & $+0.065$ & $-0.160$ & $+0.920$ \\
2745049530295263232 & 0.000589823 & $+05.380517265$ & 13.98 & 13.43 & 13.03 & 13.51 & 12.9555 & 13.5889 & 12.1934 & $+0.038$ & $-0.235$ & $+0.149$ \\
\hline
\end{tabular}
\end{table*}

\begin{figure}
\includegraphics[width=\columnwidth]{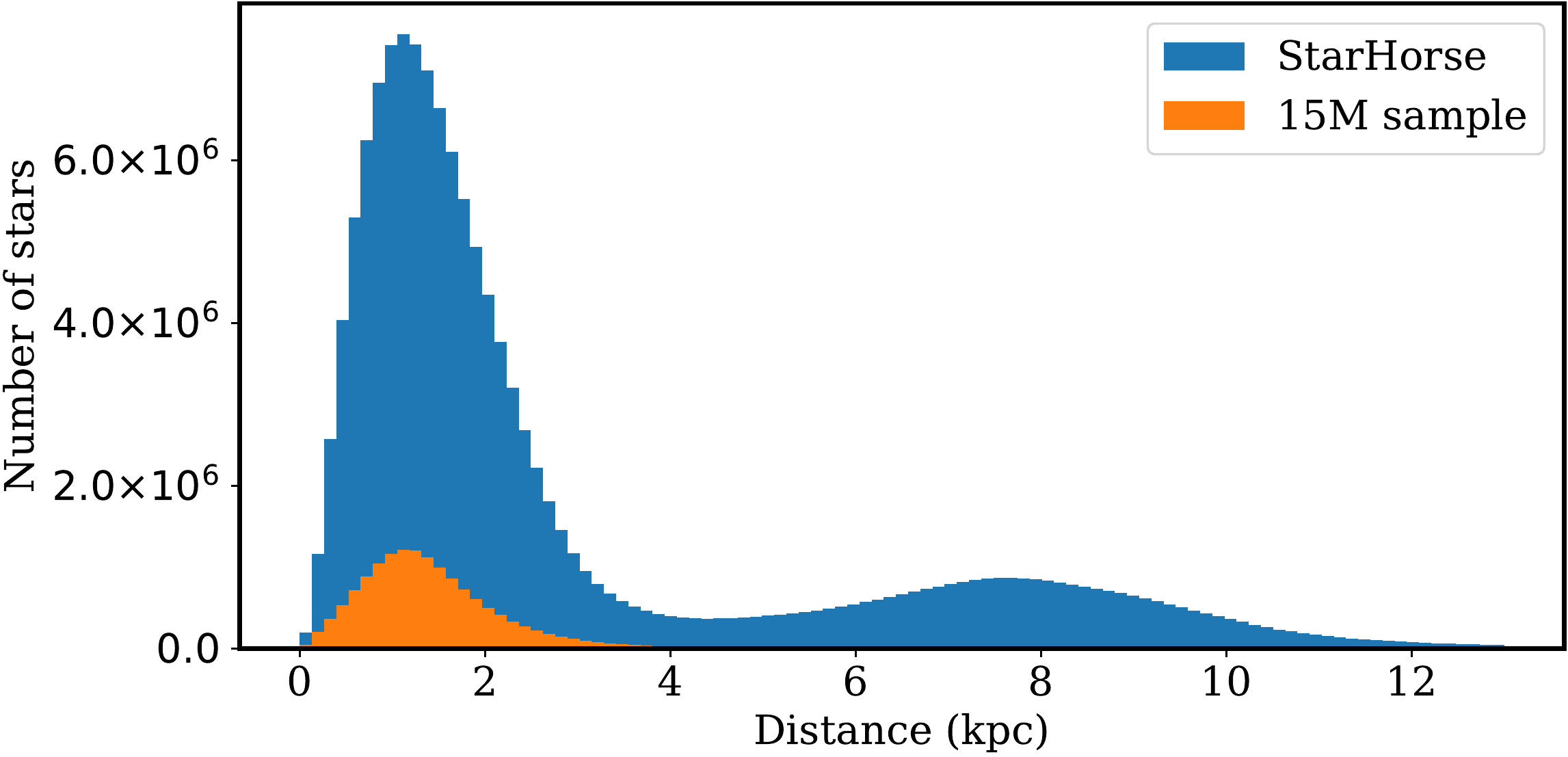}
\\[4pt]
\includegraphics[width=\columnwidth]{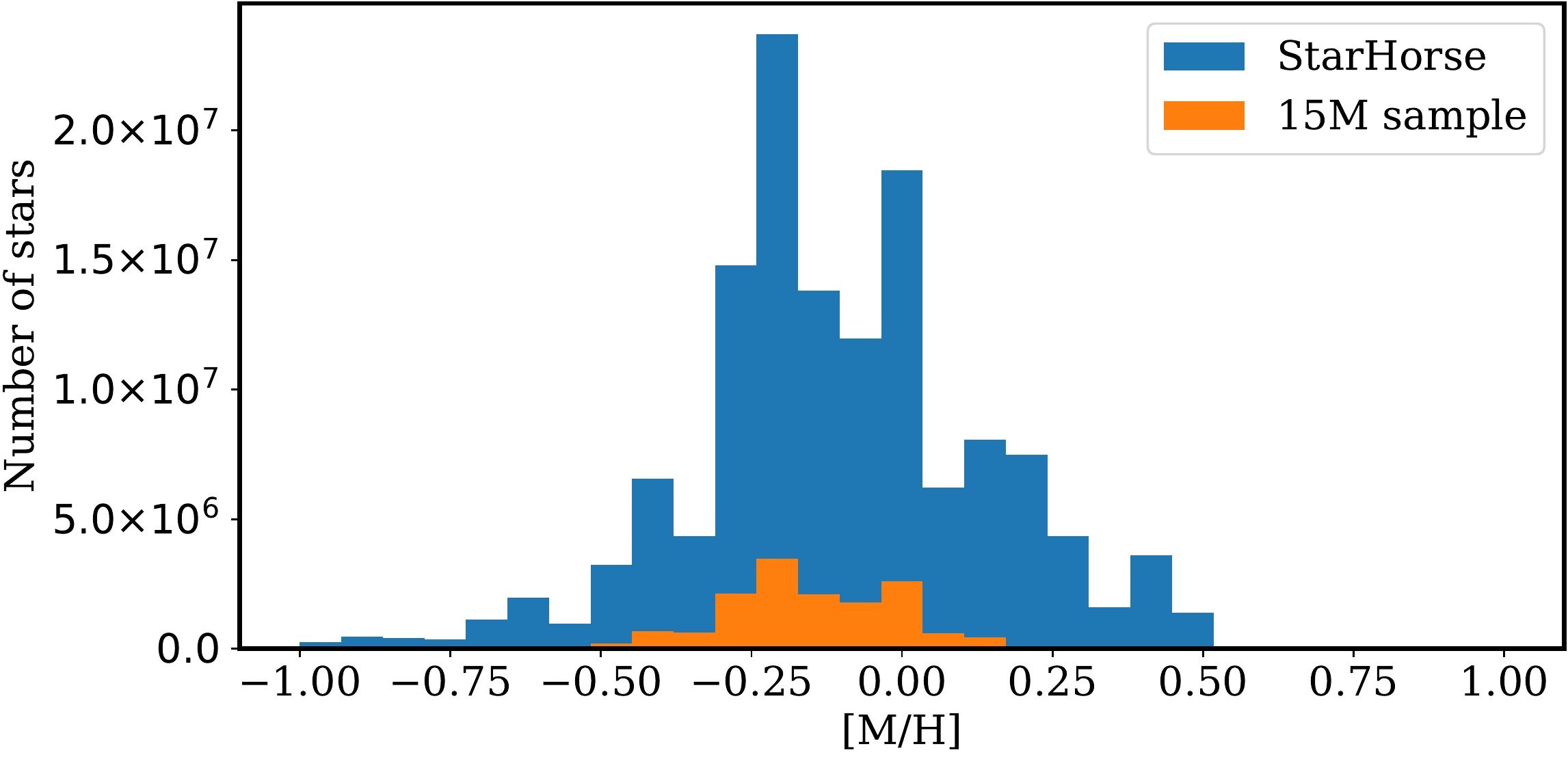}
\\[4pt]
\includegraphics[width=\columnwidth]{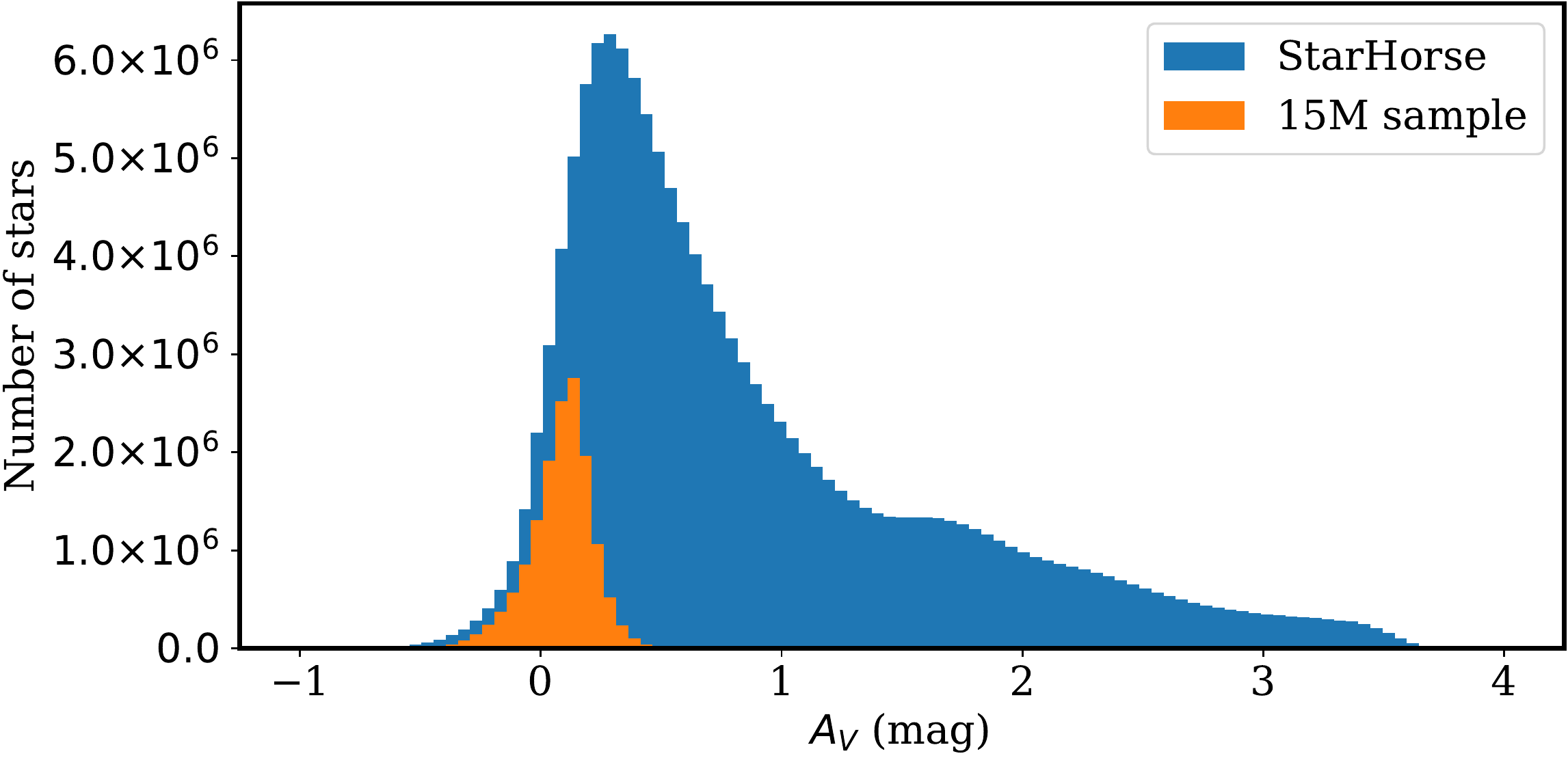}
\caption{Histograms displaying the distributions in distance (top panel),
metallicity (middle panel) and $A_{V}$ extinction (bottom panel) for the 
\mbox{136\,606\,075~stars} composing the \texttt{StarHorse} sample restricted
to \texttt{SH\_GAIAFLAG='000'} and \texttt{SH\_OUTFLAG='00000'} (blue colour),
and for the \mbox{14\,854\,959~stars} constituting the filtered 15M~star
sample (orange colour).}
\label{fig:hist_15M_SH}
\end{figure}

\begin{figure}
\includegraphics[width=\columnwidth]{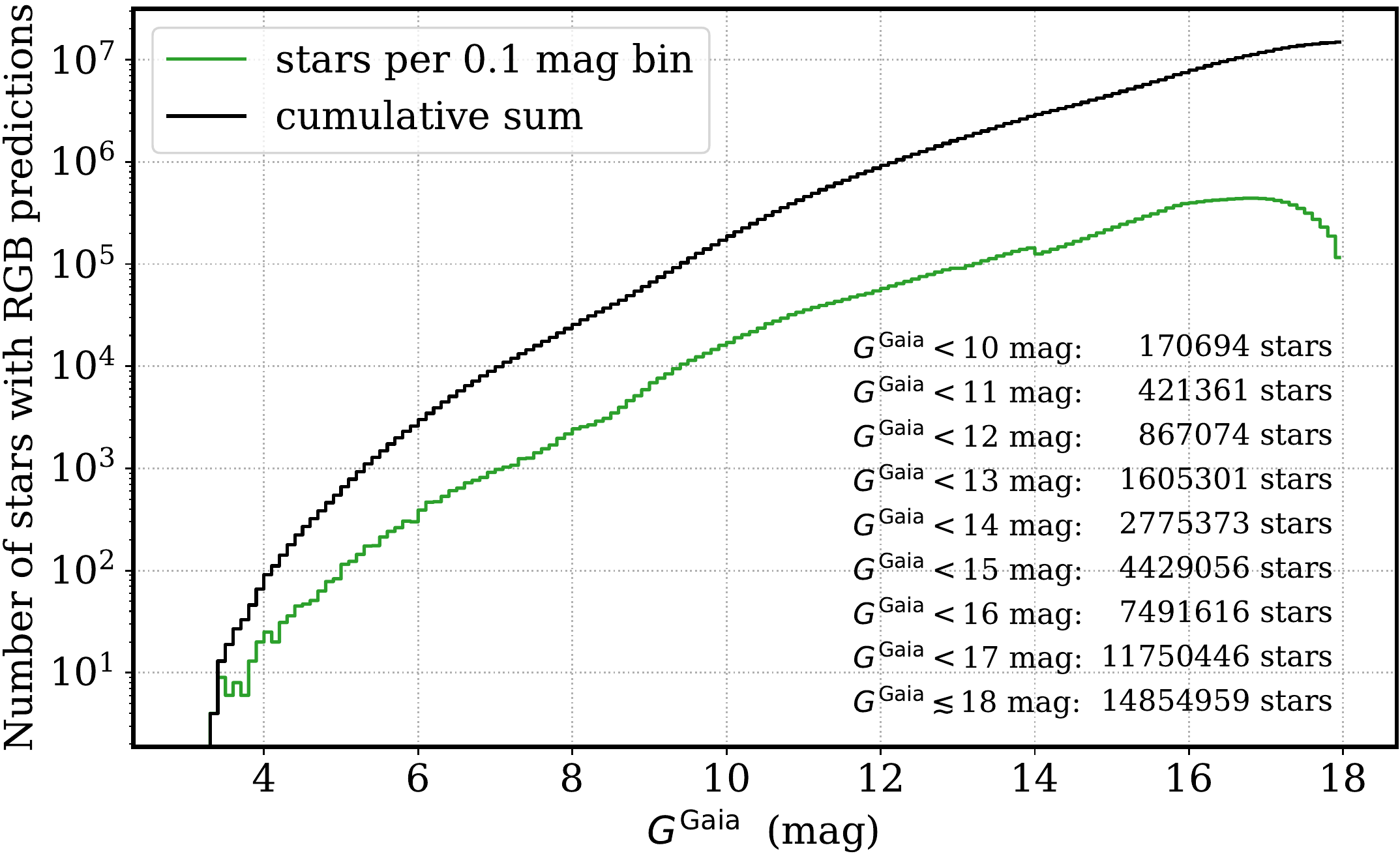}
\caption{Variation of the number of stars with RGB magnitude predictions as a
function of $G^{\rm Gaia}$. The green curve shows the histogram in bins of
0.1~mag, whereas the black line displays the cumulative sum. The inset
tabulates the cumulative sum down to some particular $G^{\rm Gaia}$
values.}
\label{fig:hist_gaia_G_cumsum}
\end{figure}

We have applied the RGB calibration derived in the
previous section to the \texttt{StarHorse} star sample of
\citet{2019A&A...628A..94A}. The data was retrieved from the \emph{Gaia}
archive hosted by the Leibniz-Institute for Astrophysics
Potsdam\footnote{\url{https://gaia.aip.de/}}. With the aim of using the star
subsample with the most reliable distance, extinction and astrophysical
parameter determinations, we restricted the initial list to
\mbox{136\,606\,075} stars flagged with \texttt{SH\_GAIAFLAG='000'} (indicating
good astrometric and photometric quality of the \emph{Gaia} DR2 data) and
\texttt{SH\_OUTFLAG='00000'} (associated to stars with reliable
\texttt{StarHorse} output parameters). It is worth noting that by adopting
these values of \texttt{SH\_GAIAFLAG} and \texttt{SH\_OUTFLAG} we excluded from
the beginning stars identified as variable in \emph{Gaia} DR2
(entry \texttt{phot\_variable\_flag} in the database), as well as objects with
significantly negative extinctions or with large $A_{V}$ uncertainties.  From
this initial collection, \mbox{13\,756\,448} stars were removed due to multiple
potential candidates in the crossmatch between DR2 and EDR3 (parameter
\texttt{dup\_max\_number} greater than one). Then, we selected stars with
extinction estimates compatible with zero within the 16th and 84th $A_{V}$
percentiles computed by \citet{2019A&A...628A..94A} when
deriving the astrophysical parameters of the \texttt{StarHorse} sample, which
restricted the sample to \mbox{20\,477\,474}~stars. At this point, the cut
\mbox{$-0.5 < G_{\rm BP}^{\rm Gaia}\!-\!G_{\rm RP}^{\rm Gaia} < 2.0$ mag} was
imposed to match the valid colour interval where
Eqs.~(\ref{eq:polyB})--(\ref{eq:polyX}) can be employed, leading to a
collection of \mbox{19\,567\,621} objects. Finally, we also constrained the
estimated metallicity for each star to be compatible with the median value
${\rm [M/H]}=-0.16$ exhibited by the C21 calibrating sample (middle panel in
Fig.~\ref{fig:hist_C21_SH}) within the 16th and 84th [M/H] percentile interval
derived by \citet{2019A&A...628A..94A}.
This last step led to the final set of \mbox{14\,854\,959} stars (hereafter the
15M star sample) for which we have estimated the RGB magnitudes.  The predicted
values are given in Table~\ref{tab:rgb_results}. 

The histograms displayed in Fig.~\ref{fig:hist_15M_SH} compare the
distributions in distance (top panel), metallicity [M/H] (middle panel) and
interstellar extinction $A_{V}$ (bottom panel) of the 15M~star subsample
(orange colour) with those exhibited by the initial $\sim 136$~million stars
corresponding to the \texttt{StarHorse} sample (blue colour). Although the
strong constraint introduced when selecting stars with extinction estimates
compatible with zero does not introduce a systematic bias in the metallicity
distribution, it imposes a clear restriction in the distance coverage: most
of the 15M~star sample is composed of stars closer than 3~kpc to the Sun.

\begin{figure*}
\includegraphics[width=0.94\columnwidth]{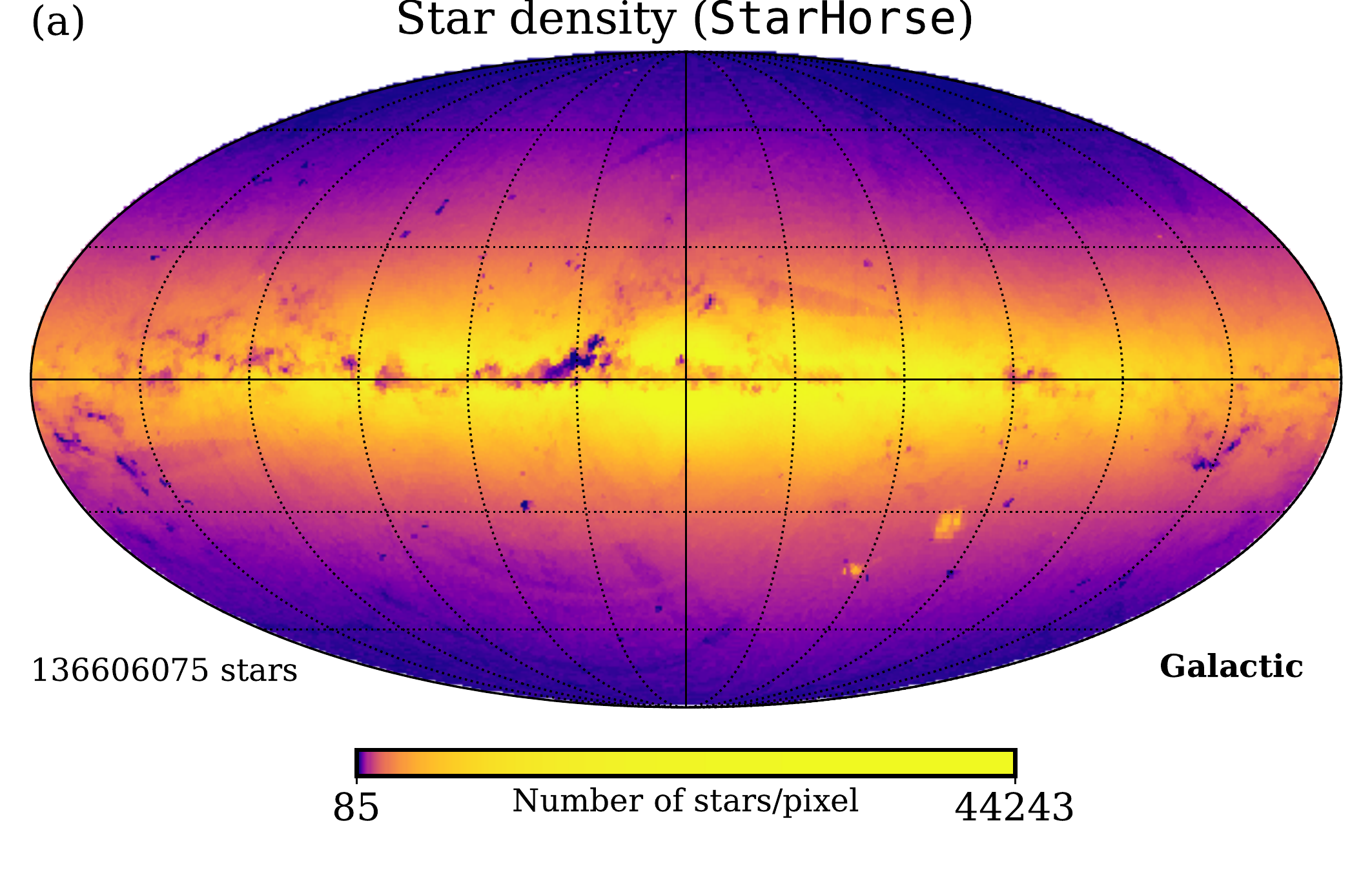}
\hfill
\includegraphics[width=0.94\columnwidth]{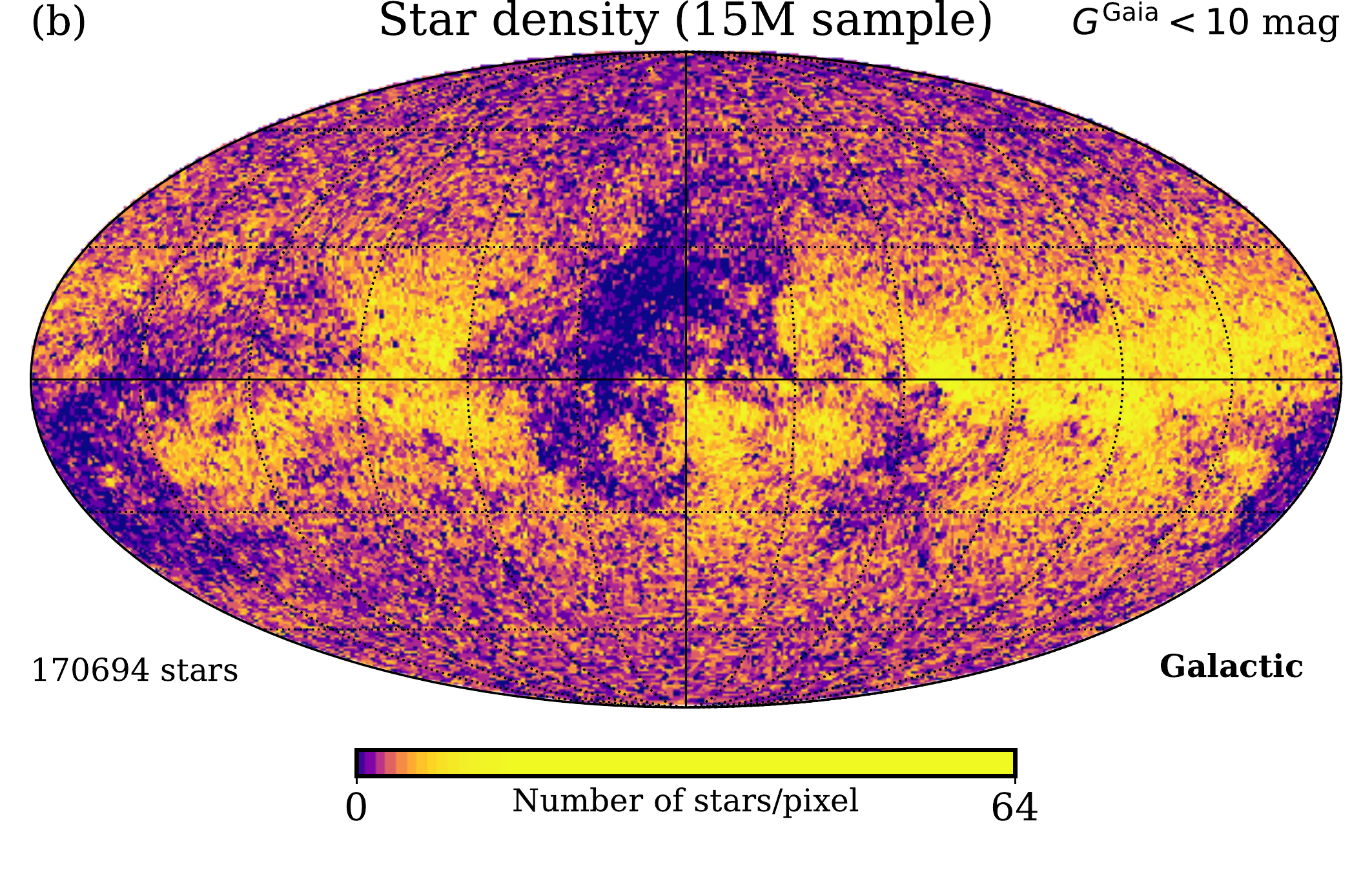}
\vskip 1mm
\includegraphics[width=0.94\columnwidth]{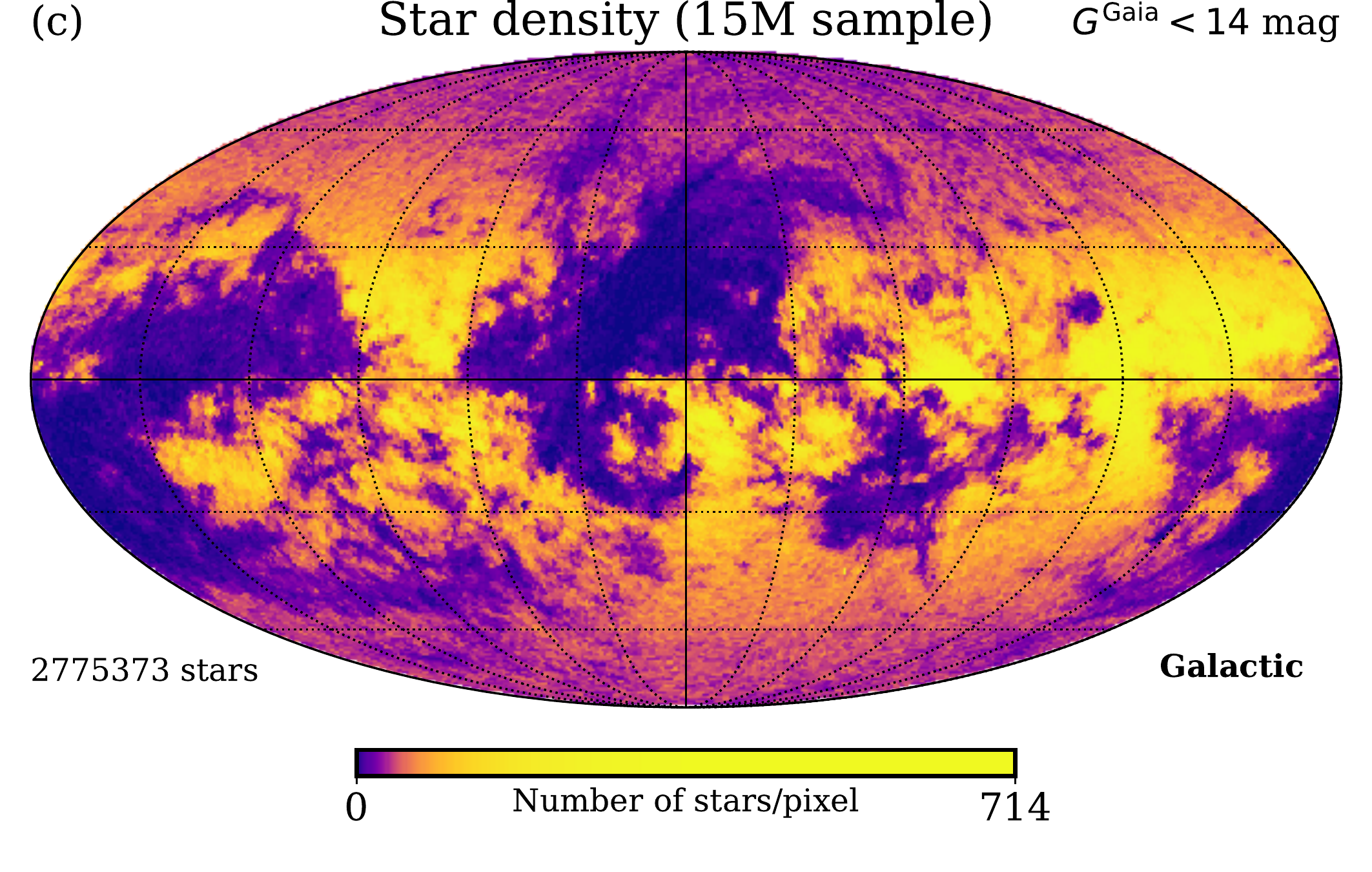}
\hfill
\includegraphics[width=0.94\columnwidth]{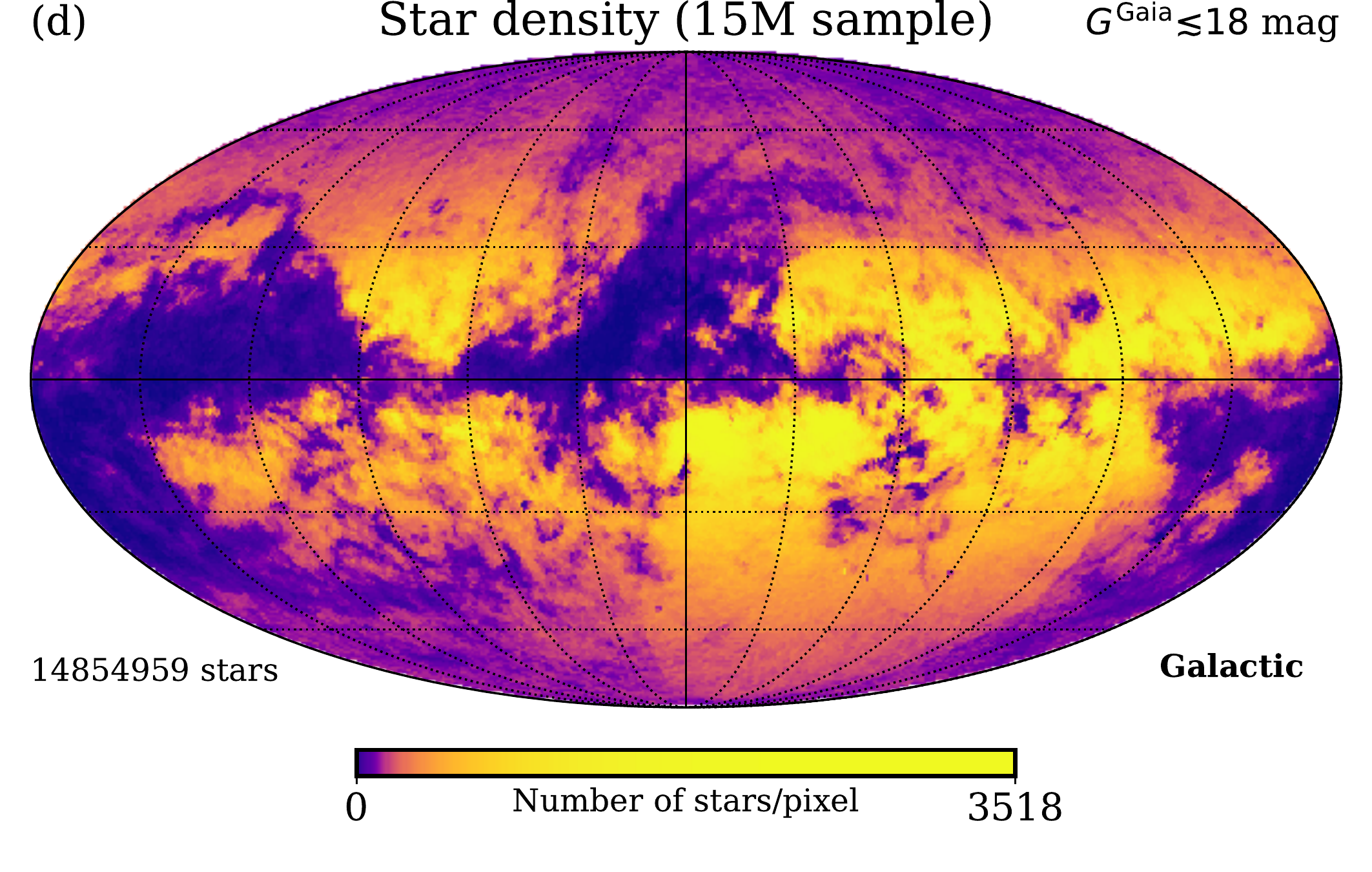}
\vskip 3mm
\includegraphics[width=0.94\columnwidth]{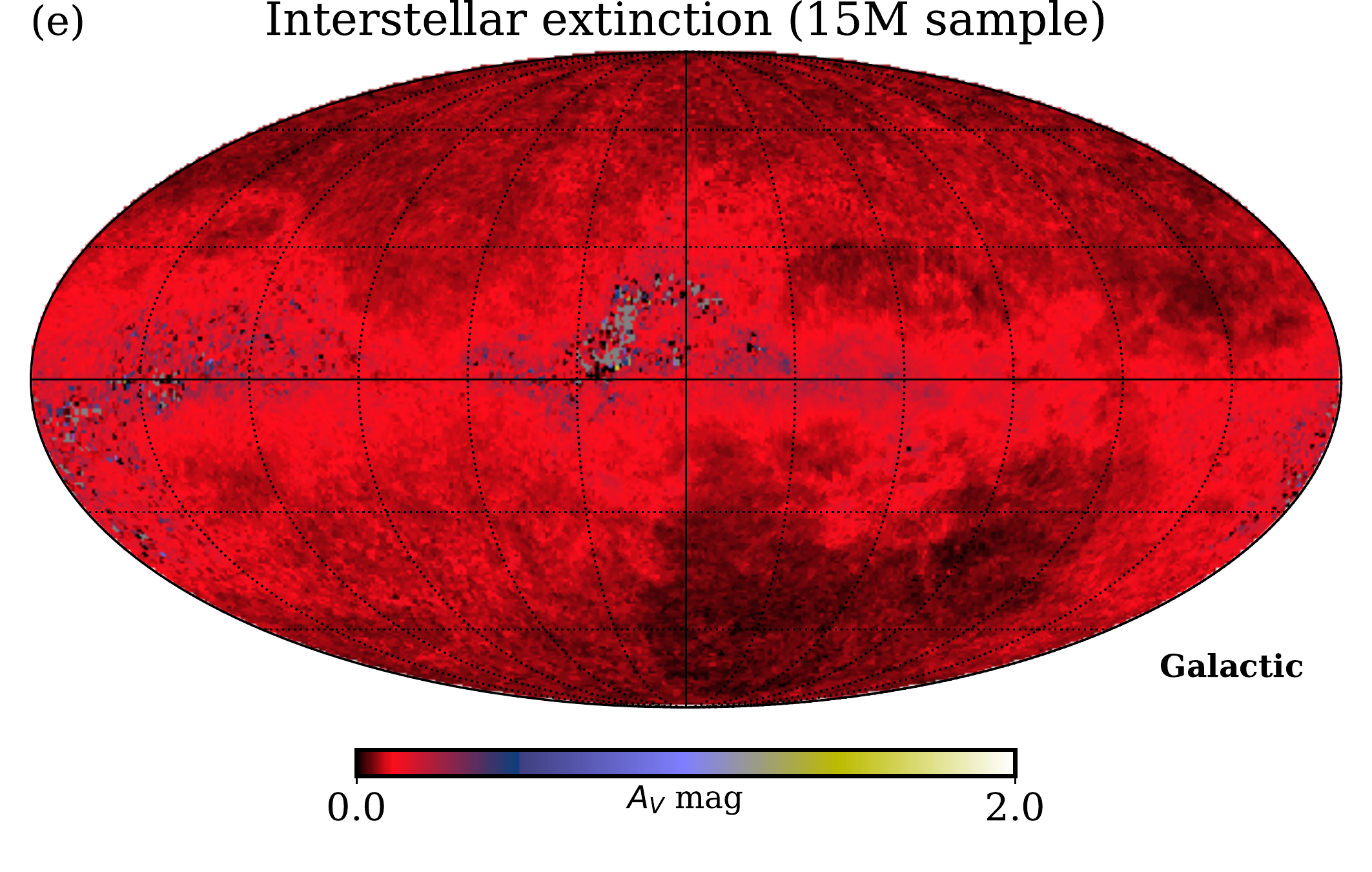}
\hfill
\includegraphics[width=0.94\columnwidth]{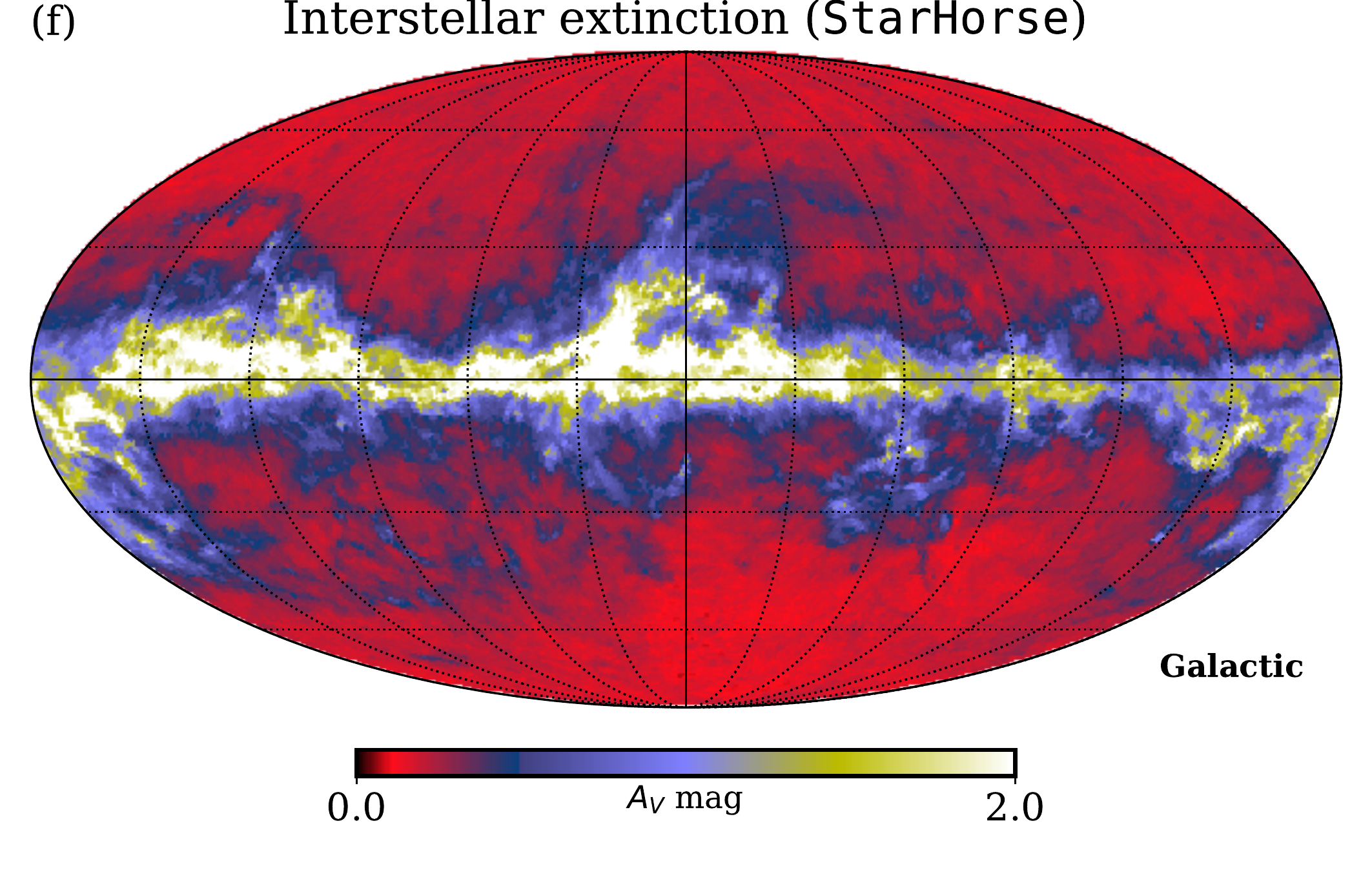}
\vskip 3mm
\includegraphics[width=0.94\columnwidth]{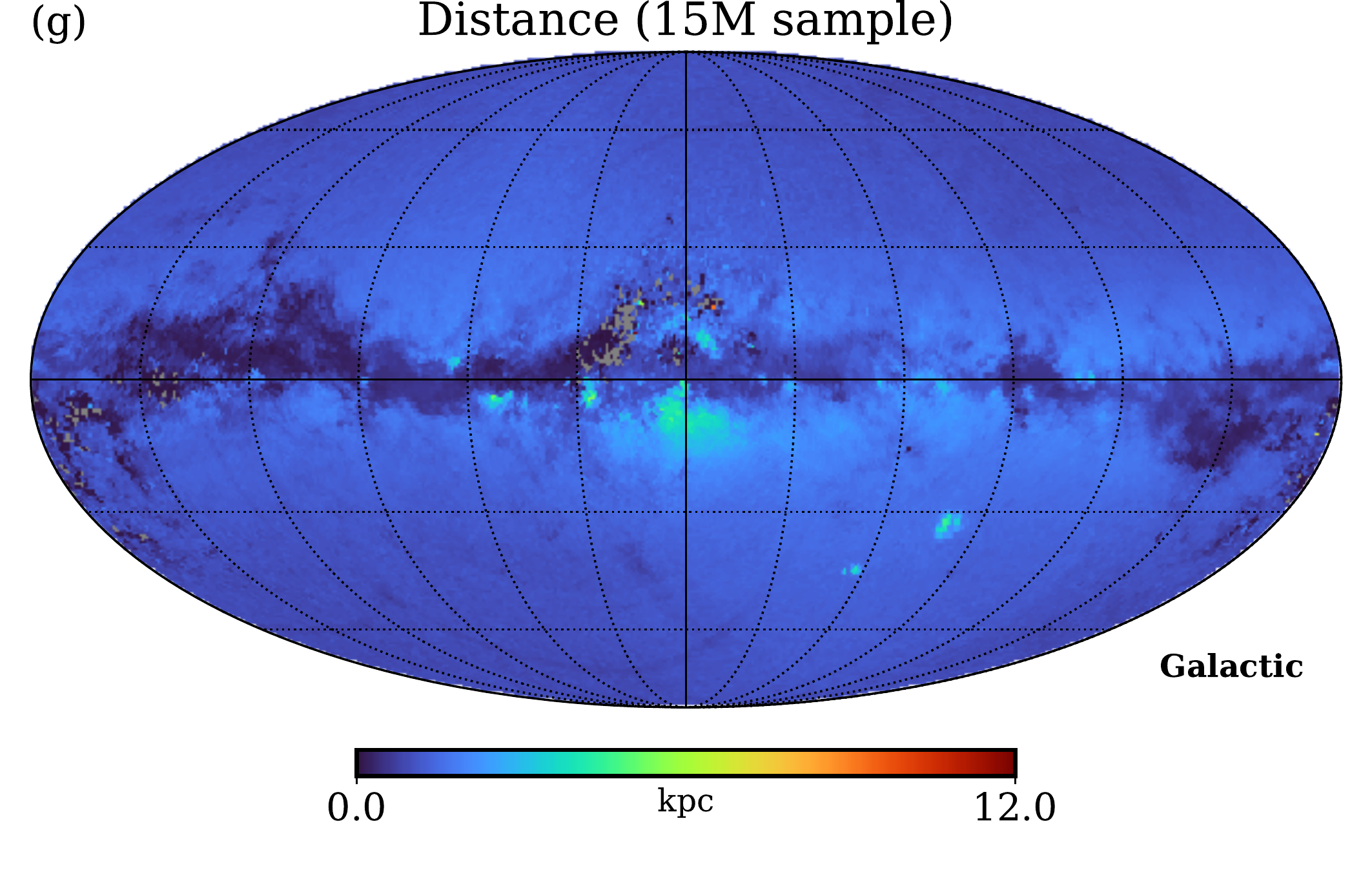}
\hfill
\includegraphics[width=0.94\columnwidth]{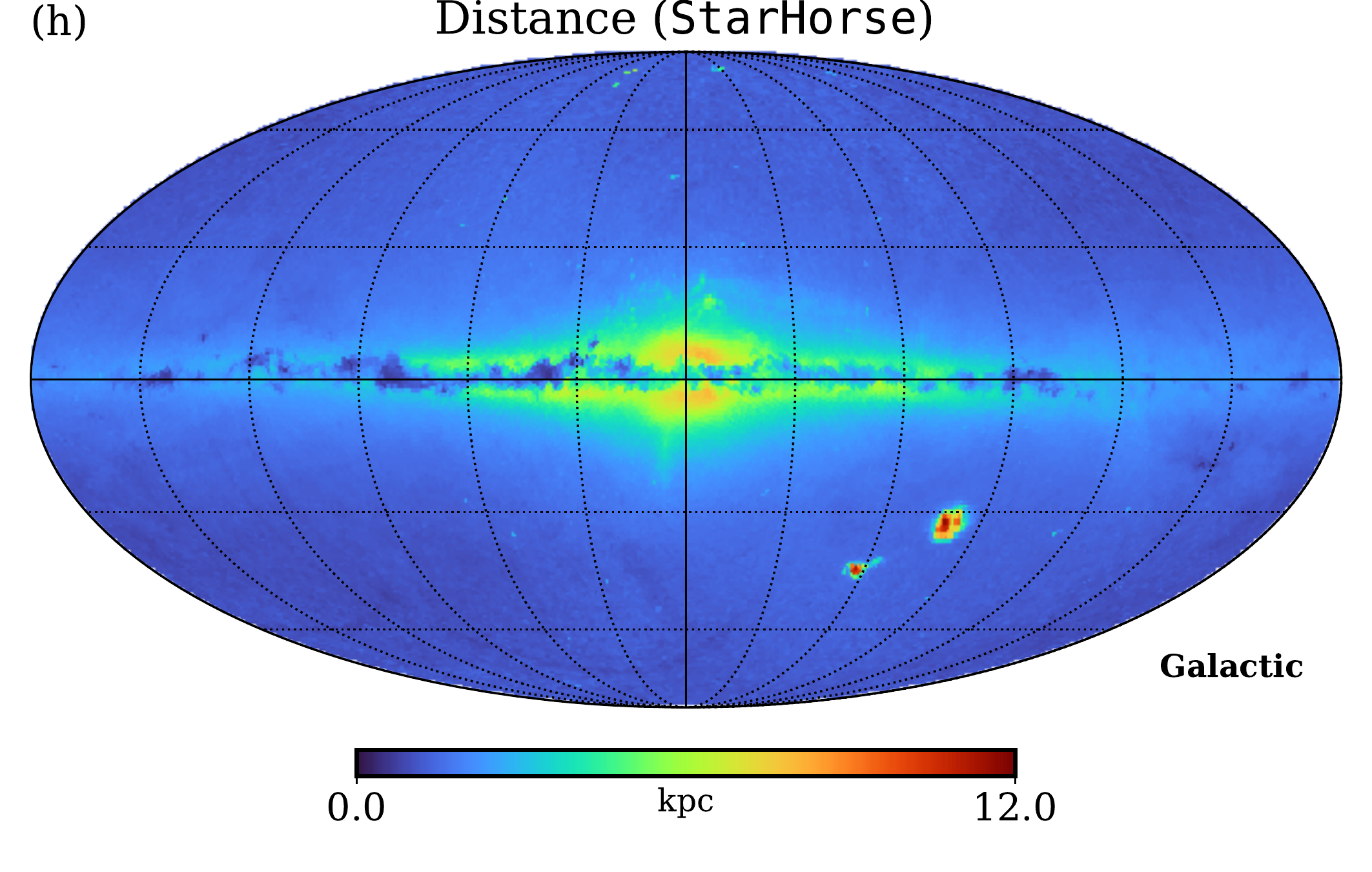}
\caption{\emph{Panel~(a):} star density map, in Galactic coordinates,
corresponding to the \mbox{136\,606\,075~stars} in the \texttt{StarHorse}
sample of \citet{2019A&A...628A..94A} flagged with \texttt{SH\_GAIAFLAG='000'}
and \texttt{SH\_OUTFLAG='00000'}. \emph{Panels~(b) (c) and~(d):} star density
maps corresponding to the stars with estimated RGB magnitudes (i.e., those in
the 15M star sample) for different limiting $G^{\rm Gaia}$ magnitudes: 10, 14
and 18~mag for panels (b), (c) and~(d), respectively. These maps have been
created using HEALPix of level~6 (providing a pixel size of 0.84 square
degrees), and are colour coded depending on the number of stars within the
pixel.  \emph{Panels~(e) and~(f):} extinction maps representing the mean
$A_{V}$ values within each HEALPix of level~6, exhibited by the 15M sample,
panel~(e), and the \mbox{136\,606\,075~stars} in the \texttt{StarHorse} sample,
panel~(f). \emph{Panels~(g) and~(h):} mean distance (kpc) within each HEALPix
of level~6 for the stars in the 15M sample, panel~(g), and the mentioned
\texttt{StarHorse} sample, panel~(h). The coordinate origin and
grid in all the panels are the same employed in
Fig.~\ref{fig:mollview_c21_bss}.}
\label{fig:star_density}
\end{figure*}

The number of stars in the 15M~sample, as a function of the $G^{\rm Gaia}$
magnitude, is shown in Fig.~\ref{fig:hist_gaia_G_cumsum} (green line). As
expected, this number increases when moving to fainter objects. The cumulative
number of stars down to a given magnitude is also displayed (black line), with
the total number of stars down to some particular $G^{\rm gaia}$ values listed
within the figure. 

The global distribution in the celestial sphere of the
\mbox{136\,606\,075~stars} in the \texttt{StarHorse} sample of
\citet{2019A&A...628A..94A} flagged with \texttt{SH\_GAIAFLAG='000'} and
\texttt{SH\_OUTFLAG='00000'} is shown in panel~(a) of
Fig.~\ref{fig:star_density} (using Galactic coordinates, with
HEALPix\footnote{\url{https://healpix.sourceforge.io/}} of level~6). It is
clear that most of the stars concentrate towards the Galactic plane. For
comparison, panels~(b), (c) and (d)~of Fig.~\ref{fig:star_density} show the
same map for the 15M star sample restricted to different limiting $G^{\rm
gaia}$ magnitudes. In addition, panels~(e) and~(f), on one hand, and panels~(g)
and~(h), on the other, compare, respectively, the mean extinction $A_{V}$ (mag)
and distance (kpc) exhibited by the 15M sample and the initial
\texttt{StarHorse} star collection.  These maps show that the stars in the 15M
sample are unevenly spread because high-extinction regions, clearly seen in
panel~(f), have been purposely avoided, being the $A_{V}$ values of the 15M
star sample, panel~(e), quite small. This behaviour is consistent with the
distance maps, which confirm that the 15M sample, panel~(g), is constituted by
nearby stars, as already shown in the histogram of
Fig.~\ref{fig:hist_15M_SH}(a), specially at low Galactic latitudes.


\section{Final discussion and conclusions}
\label{sec:conclusions}

The work presented in the previous sections has shown that it is possible to
derive simple mathematical transformations between \emph{Gaia} EDR3 photometric
data and RGB magnitudes, and to employ them to substantially extend, both in
quantity and magnitude coverage, the number of stars in the celestial sphere
with RGB magnitude estimates. The comparison with measurements performed using
stellar atmosphere models (covering [M/H] from $-2.5$ to
$+0.5$, effective temperatures from 3500 to 50000~K, and $\log g$ from 0.0 to
5.0~dex) has revealed that Eqs.~(\ref{eq:polyB})--(\ref{eq:polyX}) are
expected to provide RGB estimates within a $\pm 0.1$~mag uncertainty
when imposing the colour cut \mbox{$-0.5 < G_{\rm BP}^{\rm
Gaia}\!-\!G_{\rm RP}^{\rm Gaia} < 2.0$ mag}. The predicted
magnitudes remain within the $\pm 0.1$~mag interval even when considering
interstellar extinctions in the range $A_V \in [0.0, 3.0]$~mag, although there
are some systematic deviations depending on the $G_{\rm BP}^{\rm
Gaia}\!-\!G_{\rm RP}^{\rm Gaia}$ colour, being $G^{\rm rgb}$ and $G_{\rm
BR}^{\rm rgb}$ more robust to this effect than $B^{\rm rgb}$ and $R^{\rm rgb}$.
For that reason, we have restricted the star sample to nearby objects for which
the interstellar extinction estimate derived by \citet{2019A&A...628A..94A} was
statistically compatible with zero.

The new catalogue, encompassing \mbox{$\sim15$~million} stars, should smooth 
the way for:
\begin{itemize}
\item[i)] The use of the standard RGB photometric system proposed by C21: the
homogenization of RGB measurements, derived from data obtained with a
potentially very large number of different cameras, will become essential. This
is a critical aspect that can hardly be overestimated.

\item[ii)] The proper calibration of commercial-grade RGB cameras: this task
will be facilitated by considering the wide magnitude range exhibited by the
15M star sample, leaving ample room for the use of RGB imaging instruments with
different field of views, as well as their exploitation in scientific projects
requiring distinct exposure times.

\item[iii)] The calibration of observations carried out in any region of the
sky, even at high Galactic latitudes. Note, however, that the distribution in
the celestial sphere is inhomogeneous and depends on the interstellar
extinction in the direction of observation, being the total number of available
calibrated stars unavoidably tied to the adopted limiting magnitude.

\item[iv)] The correction of the measurements for atmospheric extinction: this
data reduction step is particularly important in wide-field exposures, where
stars at different airmasses are simultaneously observed.
\end{itemize}

It is important to highlight that the RGB magnitude predictions computed for
the 15M star sample should not be considered to be extremely accurate on a star
by star basis. In particular, although stars detected as variables in DR2 have
been removed, it is likely that variable sources are still
hidden in the selected sample. In any case, the large number of calibrating
stars available should facilitate the computation of statistical averages that
allow the rejection of potential outliers, guaranteeing adequate calibrations.

Although in this work we have restricted the RGB estimates to a subsample of
the stars published by \citet{2019A&A...628A..94A} with good photo-astrometric
distances, extinctions and astrophysical parameters, it is possible to apply
the derived calibrations to many more \emph{Gaia} EDR3 stars for which these
parameters are not even available. In principle, this should only be employed
for stars with high Galactic latitude in order to minimize systematic errors
introduced by interstellar extinction (although there are regions with low
galactic latitude and low extinction), and within the colour interval
\mbox{$-0.5 < G_{\rm BP}^{\rm Gaia}\!-\!G_{\rm RP}^{\rm Gaia} < 2.0$ mag}. In
this regard, it is interesting to note that $\sim 82$~per cent of the
\mbox{304\,602\,695} stars available in \emph{Gaia} EDR3 with \mbox{$G^{\rm
Gaia} \leq 18$~mag} \mbox{(251\,118\,359 stars)} verify the last colour cut.
This means that, down to a given limiting magnitude, the total
number of observable stars is expected to be much larger than the number of
stars belonging to the 15M~star sample. The initial calibration obtained by
employing the predicted RGB magnitudes presented here, using only a subsample
of the observed stars, can be applied, in a second iteration, to the remaining
stars verifying the appropriate colour cut. The inclusion of additional stars
(after removing outliers) should facilitate the computation of a more reliable
calibration. In addition, the relative robustness of $G^{\rm rgb}$ and $G_{\rm
BR}^{\rm rgb}$ to moderate amounts of interstellar extinction, should even
allow the use of observations performed in regions closer to the Galactic plane
(although this should always be double-checked through comparison with
calibrations performed at higher Galactic latitudes). With the aim of helping
on the use (and extension) of the 15M star sample, in
Appendix~\ref{ap:beyond_15M} we illustrate how to estimate RGB magnitudes for
all the \emph{Gaia} EDR3 stars within a particular cone search. 


The synthetic RGB photometry presented here fills an important gap that can
help to provide a firm ground for accurate camera calibrations and the
systematic exploitation of RGB photometry.


\section*{Acknowledgements}

The authors are grateful for the careful reading by the
referee, whose constructive remarks have helped to improve the paper, making
the text more precise and readable.
The authors acknowledge financial support from the Spanish Programa Estatal de
I+D+i Orientada a los Retos de la Sociedad under grant RTI2018-096188-B-I00,
which is partly funded by the European Regional Development Fund (ERDF),
S2018/NMT-4291 (TEC2SPACE-CM), and ACTION, a project funded by the European
Union H2020-SwafS-2018-1-824603. The participation of ICCUB researchers was
(partially) supported by the Spanish Ministry of Science, Innovation and
University (MICIU/FEDER, UE) through grant RTI2018-095076-B-C21, and the
Institute of Cosmos Sciences University of Barcelona (ICCUB, Unidad de
Excelencia ’Mar\'{\i}a de Maeztu’) through grant CEX2019-000918-M.  SB
acknowledges Xunta de Galicia for financial support under grant ED431B 2020/29.
The participation of ASdM was (partially) supported by the EMISSI@N project
(NERC grant NE/P01156X/1). This work has made use of data from the European
Space Agency (ESA) mission {\it Gaia} (\url{https://www.cosmos.esa.int/gaia}),
processed by the {\it Gaia} Data Processing and Analysis Consortium (DPAC,
\url{https://www.cosmos.esa.int/web/gaia/dpac/consortium}). Funding for the
DPAC has been provided by national institutions, in particular the institutions
participating in the {\it Gaia} Multilateral Agreement.  This work has been
possible thanks to the extensive use of IPython and Jupyter notebooks
\citep{PER-GRA:2007}, as well as the software package
R\footnote{\url{https://www.R-project.org/}} \citep{Rcitation}. This
research made use of {\sc astropy}\footnote{\url{http://www.astropy.org}}, a
community-developed core Python package for Astronomy \citep{astropy:2013,
astropy:2018}, {\sc numpy} \citep{harris2020array}, {\sc scipy} \citep{2020SciPy-NMeth},
{\sc matplotlib} \citep{4160265}, {\sc pandas}
\citep{mckinney-proc-scipy-2010}, and {\sc vaex}
\citep{2018A&A...618A..13B}. Some of the results in this paper have been
derived using the healpy and HEALPix packages \citep{2005ApJ...622..759G,
Zonca2019}. This research has made use of the Simbad database and the VizieR
catalogue access tool, CDS, Strasbourg, France \mbox{(DOI:
10.26093/cds/vizier)}. The original description of the VizieR service was
published in \mbox{A\&AS 143, 23}.

\section*{Data Availability}

The work in this paper has made use of \emph{Gaia} DR2 and EDR3, provided by
the European Space Agency\footnote{\url{https://gea.esac.esa.int/archive/}},
the \texttt{StarHorse} database hosted by the Leibniz-Institute for
Astrophysics
Potsdam\footnote{\url{https://data.aip.de/projects/starhorse2019.html}}, the
Stellar Atmopshere Models of \citet{2003IAUS..210P.A20C}, as provided by the
STScI web
page\footnote{\url{https://www.stsci.edu/hst/instrumentation/reference-data-for-calibration-and-tools/astronomical-catalogs/castelli-and-kurucz-atlas}},
the Bright Star Catalogue \citep[][available
online\footnote{\url{https://vizier.u-strasbg.fr/viz-bin/VizieR-3?-source=V/50/catalog}}]{1964cbs..book.....H},
and the catalogue of synthethic RGB magnitudes published by
\citet{2021MNRAS.504.3730C}.

All the results of this paper, together with future additional material, is
available online at \url{http://guaix.fis.ucm.es/~ncl/rgbphot/gaia},
and will be also available through VizieR.



\bibliographystyle{mnras}
\bibliography{paper_RGB_Gaia}




\appendix
\section{Estimation of RGB magnitudes beyond the 15~million star sample}
\label{ap:beyond_15M}

We illustrate in this appendix how to estimate RGB magnitudes for all the
\emph{Gaia} EDR3 stars within an arbitrary cone search in the celestial sphere.
Their use in any calibration procedure should always be accompanied by the
observation of reference stars belonging to the 15M star sample.

It is important to remember that \emph{Gaia} has a bright limit around $G^{\rm
Gaia} \sim 3$~mag, and thus, bright stars will be missing.

\subsection{Use of an ADQL query}
\label{ap:adql_query}

\begin{figure}
\includegraphics[width=\columnwidth]{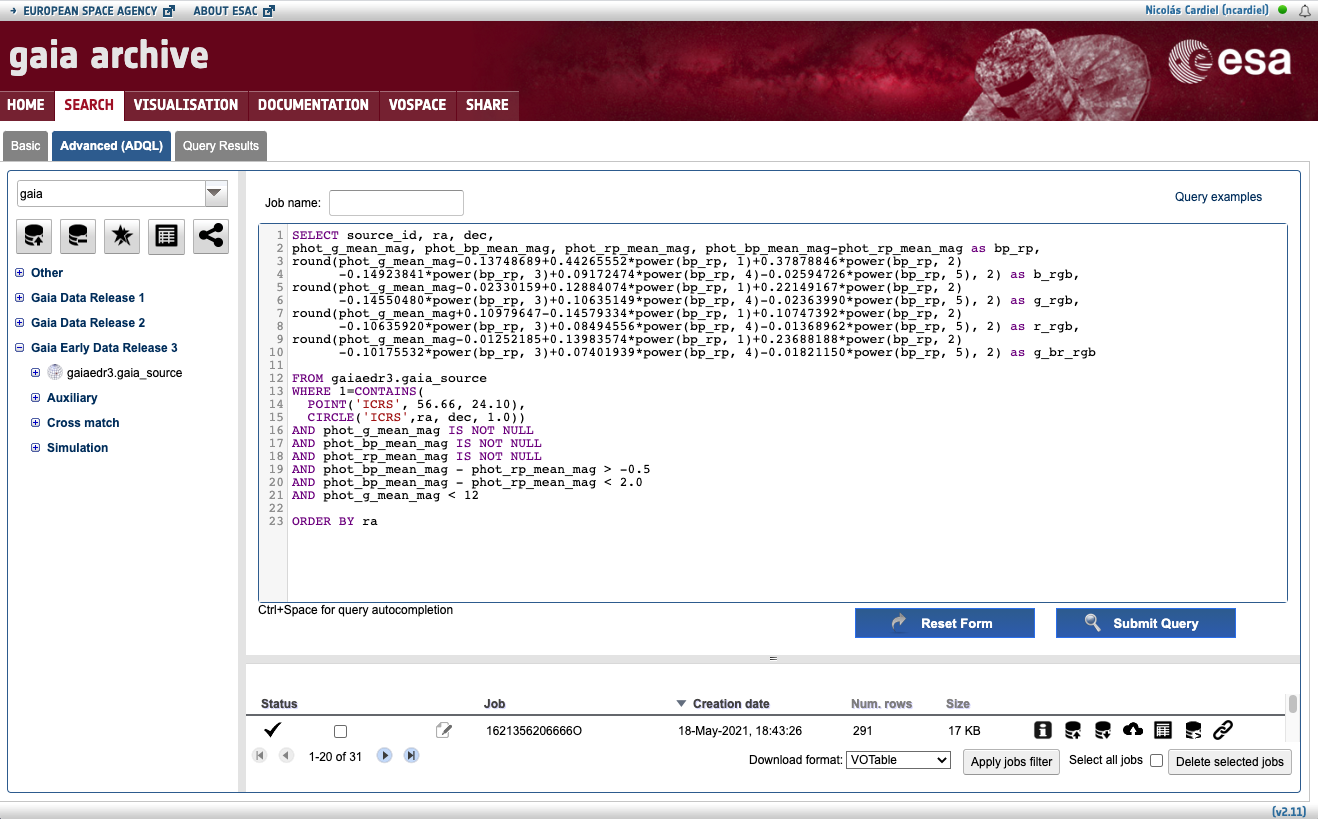}

\vspace{2mm}

\includegraphics[width=\columnwidth]{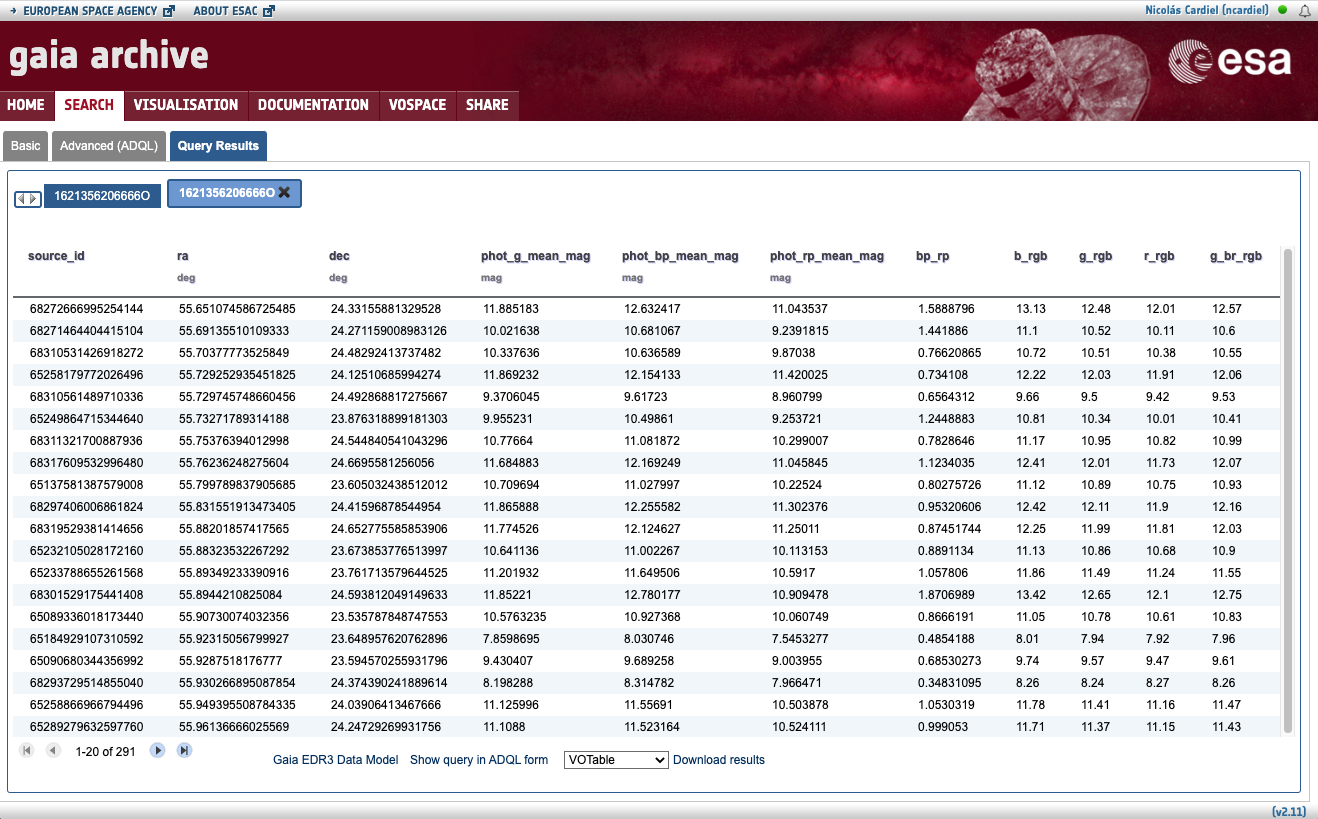}
\caption{Screenshots of the results of using the ADQL query described in
Sect.~\ref{ap:adql_query} in the European Agency Portal of the \textit{Gaia}
EDR3 database. \textit{Top panel}: interface for advanced queries. The cone
search is centrered at RA=56.66\degr, DEC=24.10\degr\ (line number~14 of the
displayed query), with a search radius of 1\degr\ (line~15), and a limiting
$G^{\rm Gaia}$ magnitude of 12~mag (line~21). \textit{Bottom panel}: table with
results. A total of 291 stars match the selection criteria (only the first 20
are displayed), with the RGB magnitude estimates $B^{\rm rgb}$, $G^{\rm rgb}$,
$R^{\rm rgb}$ and $G_{\rm BR}^{\rm rgb}$ listed in the last four columns,
respectively.}
\label{fig:gaia_query}
\end{figure}

A direct way to retrieve RGB magnitudes is to employ an ADQL
\citep[Astronomical Data Query Language; see e.g.][]{2008ivoa.spec.1030O} query
to access the \emph{Gaia} EDR3 database. This language allows to evaluate
mathematical expressions on the relevant parameters.  For illustration, a
simple cone search of \emph{Gaia} EDR3 stars brighter than \mbox{$G^{\rm
Gaia}\!=\!12$~mag}, within a circular region of radius 1\degr, with centre at
RA=56.66\degr and DEC=24.10\degr\ (corresponding to the Pleiades star cluster),
imposing the colour selection \mbox{$-0.5 < G_{\rm BP}^{\rm Gaia}\!-\!G_{\rm
RP}^{\rm Gaia} < 2.0 $ mag}, and using the polynomial functions given
Eqs.~(\ref{eq:polyB})--(\ref{eq:polyX}) to estimate the RGB magnitudes, can be
performed employing:

\begin{verbatim}
SELECT source_id, ra, dec,
phot_g_mean_mag, phot_bp_mean_mag, phot_rp_mean_mag,
phot_bp_mean_mag-phot_rp_mean_mag as bp_rp,
round(phot_g_mean_mag 
      -0.13748689
      +0.44265552*power(bp_rp, 1)
      +0.37878846*power(bp_rp, 2)
      -0.14923841*power(bp_rp, 3)
      +0.09172474*power(bp_rp, 4)
      -0.02594726*power(bp_rp, 5), 2) as b_rgb,
round(phot_g_mean_mag 
      -0.02330159
      +0.12884074*power(bp_rp, 1)
      +0.22149167*power(bp_rp, 2)
      -0.14550480*power(bp_rp, 3)
      +0.10635149*power(bp_rp, 4)
      -0.02363990*power(bp_rp, 5), 2) as g_rgb,
round(phot_g_mean_mag 
      +0.10979647
      -0.14579334*power(bp_rp, 1)
      +0.10747392*power(bp_rp, 2)
      -0.10635920*power(bp_rp, 3)
      +0.08494556*power(bp_rp, 4)
      -0.01368962*power(bp_rp, 5), 2) as r_rgb,
round(phot_g_mean_mag 
      -0.01252185
      +0.13983574*power(bp_rp, 1)
      +0.23688188*power(bp_rp, 2)
      -0.10175532*power(bp_rp, 3)
      +0.07401939*power(bp_rp, 4)
      -0.01821150*power(bp_rp, 5), 2) as g_br_rgb
\end{verbatim}
\texttt{FROM gaiaedr3.gaia\_source}\\
\texttt{WHERE 1=CONTAINS(}\\
\mbox{}\texttt{\;\;\;\;POINT('ICRS', {\bf 56.66}, {\bf 24.10}),}\\
\mbox{}\texttt{\;\;\;\;CIRCLE('ICRS',ra, dec, {\bf 1.00}))}\\
\texttt{AND phot\_g\_mean\_mag IS NOT NULL}\\
\texttt{AND phot\_bp\_mean\_mag IS NOT NULL}\\
\texttt{AND phot\_rp\_mean\_mag IS NOT NULL}\\
\texttt{AND phot\_bp\_mean\_mag - phot\_rp\_mean\_mag > -0.5}\\
\texttt{AND phot\_bp\_mean\_mag - phot\_rp\_mean\_mag < 2.0}\\
\texttt{AND phot\_g\_mean\_mag < {\bf 12}}\\
\texttt{ }\\
\texttt{ORDER BY ra}\\[2pt]

Note that in the previous example the user-defined central coordinates, search
radius and limiting $G^{\rm Gaia}$ magnitude are shown in boldface.
Fig.~\ref{fig:gaia_query} illustrates the execution of this query through the
\emph{Gaia} EDR3 Archive at the European Space
Agency\footnote{\url{https://gea.esac.esa.int/archive/}}, which
returns RGB magnitude estimates for the sample of 291~stars matching the
selection criteria. Note, however, that at this point the resulting star list
should be cross-matched with the 15M star sample built in this work in
order to segregate the star sample into reference stars (those belonging to the
15M sample) and secondary calibrating stars (those that do not
belong). Since this cross-matching goes beyond the idea of the simple ADQL
query employed here, in the next subsection we describe an auxiliary Python
package that performs this task automatically.

\subsection{The Python {\sc rgblues} package}

\begin{figure*}
\includegraphics[width=\textwidth]{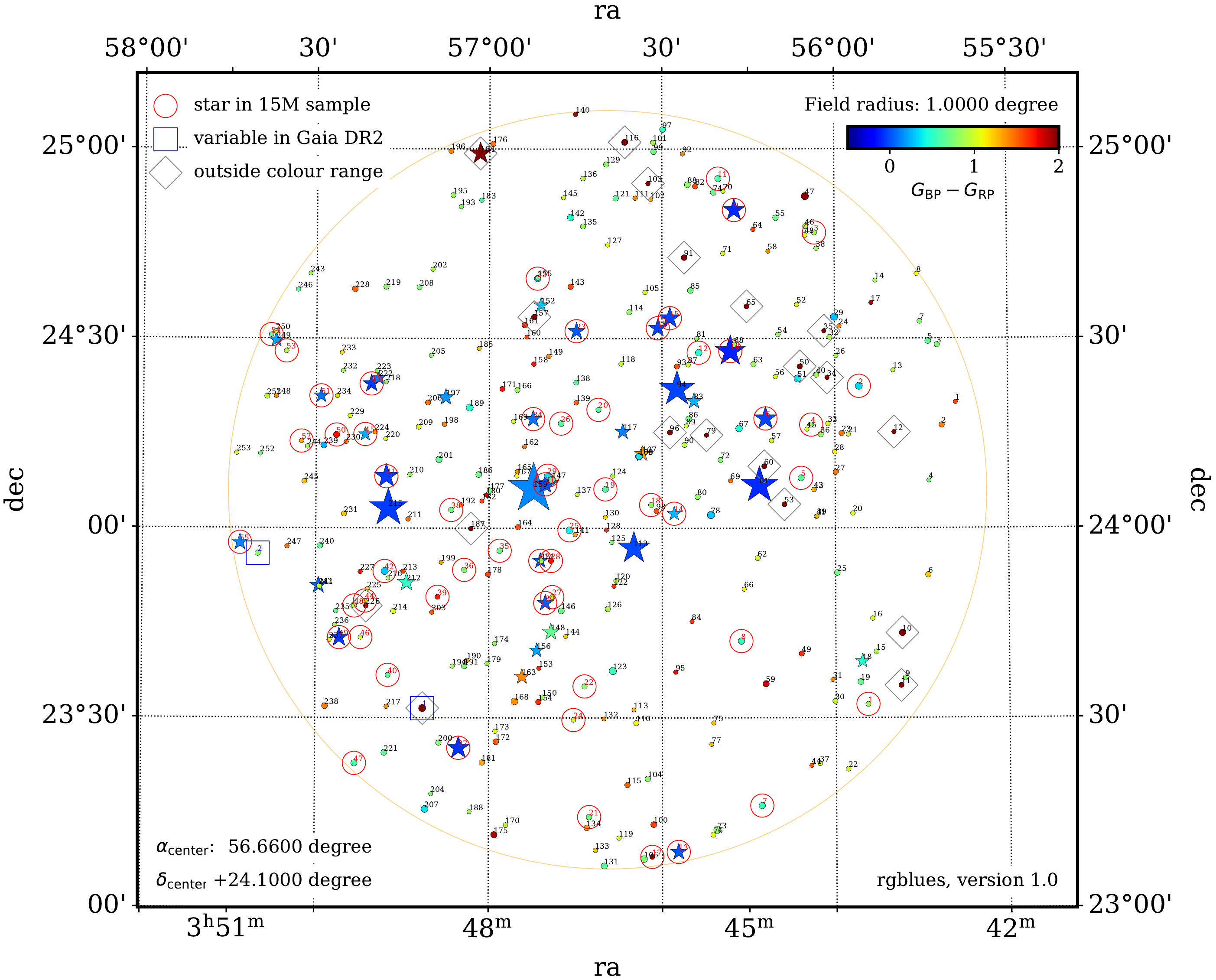}
\caption{Example of finding chart generated by the Python package
{\sc rgblues} after performing a cone search centred in the
Pleiades star cluster, with a search radius of 1\degr. The stars in this plot
are colour coded based on the \mbox{$G_{\rm BP}^{\rm Gaia}\!-\!G_{\rm RP}^{\rm
Gaia}$} colour. A red circle has been overplotted on the stars belonging to the
15M star sample, a blue square on the variable objects in DR2, and a
grey diamond on EDR3 stars outside the \mbox{$-0.5 < G_{\rm BP}^{\rm
Gaia}\!-\!G_{\rm RP}^{\rm Gaia} < 2.0$~mag} colour interval. Stars brighter
than a pre-defined threshold are displayed with big star symbols. The stars are
numbered with labels of different colours (red, blue and black for stars in the
15M star sample, variable objects in DR2, and remaining stars in EDR3,
respectively), matching the order of the stars in the three output CSV files
generated during the execution of the program.}

\label{fig:RGBfromGaiaEDR3_pleiades}
\end{figure*}

Trying to help future users of the 15M star sample to perform cone
search queries as that shown in the previous subsection, we have also created a
Python package, called {\sc rgblues}\footnote{Available at
\url{https://github.com/guaix-ucm/rgblues}}, that executes this type
of queries and performs the additional extra work required to automatically
discriminate between reference stars, belonging to the 15M star sample, from
secondary calibrating stars (additional objects in EDR3), flagging objects
detected to be variable in \textit{Gaia} DR2 and those outside the colour range
\mbox{$-0.5 < G_{\rm BP}^{\rm Gaia}\!-\!G_{\rm RP}^{\rm Gaia} < 2.0$~mag}. 

Once installed, the software can be easily executed from the command line:\\
\texttt{\$ rgblues 56.66 24.10 1.0 12}\\ 
The four positional arguments correspond to RA, DEC, search radius
(these three parameters in decimal degrees) and limiting $G^{\rm Gaia}$
magnitude. 

The steps followed by {\sc rgblues} to complete its tasks are the following:
\begin{enumerate}
\item[Step 1:] cone search in \textit{Gaia} EDR3 down to a pre-defined limiting
$G^{\rm gaia}$ magnitude, gathering the following
parameters: \texttt{source\_id}, \texttt{ra}, \texttt{dec},
\texttt{phot\_g\_mean\_mag}, \texttt{phot\_bp\_mean\_mag} and
\texttt{phot\_rp\_mean\_mag}. In this case, an ADQL query similar to that shown
in Sect.~\ref{ap:adql_query} is performed, without imposing any $G_{\rm BP}^{\rm
Gaia}\!-\!G_{\rm RP}^{\rm Gaia}$ colour restriction nor evaluating the
polynomial transformations given in Eqs.~(\ref{eq:polyB})--(\ref{eq:polyX}).

\item[Step 2:] cone search in the \texttt{StarHorse} sample through the 
\emph{Gaia} archive hosted by the Leibniz-Institute for Astrophysics
Potsdam. This step, which is optional, allows the compilation of stellar
parameters associated with each star, such as interstellar extinction,
metallicity and distance.

\item[Step 3:] cross-matching of the previous EDR3 sample with the list of 15M
star sample from this work. This step determines the subsample of EDR3
stars for which the RGB photometric calibration is reliable.

\item[Step 4:] cone search in \textit{Gaia} DR2. This additional step is
performed to retrieve the \texttt{phot\_variable\_flag} parameter
indicating whether the star was flagged as variable in DR2. Note that this flag
is not available in EDR3.

\item[Step 5:] cross-matching between DR2 and EDR3 to identify the variable
stars in EDR3. This step is required because it is not guaranteed that the same
astronomical source will always have the same source identifier in the
different \textit{Gaia} Data Releases.

\item[Step 6:] computation of the RGB magnitudes using the polynomial
transformations given in Eqs.~(\ref{eq:polyB})--(\ref{eq:polyX}).

\item[Step 7:] generation of the output files. Three files (in CSV format) are
generated, segregating the star list in i)~stars belonging to the 15M
star sample (with reliable RGB magnitude estimates), ii)~objects flagged as
variable in \textit{Gaia} DR2, and iii)~remaining objects in \textit{Gaia}
EDR3. Note that the RGB magnitudes estimated for the latter can be potentially
biased due to systematic effects introduced by interstellar extinction, or by
exhibiting non-solar metallicity or a colour outside the \mbox{$-0.5 < G_{\rm
BP}^{\rm Gaia}\!-\!G_{\rm RP}^{\rm Gaia} < 2.0$~mag} interval.

\item[Step 8:] creation of a finding chart.
Fig.~\ref{fig:RGBfromGaiaEDR3_pleiades} illustrates the resulting plot after
executing the Python code with the same cone-search parameters employed in
Sect.~\ref{ap:adql_query}.

\end{enumerate}


\bsp	
\label{lastpage}
\end{document}